\documentclass[useAMS,usenatbib]{mn2e}

\usepackage[pdftex]{graphicx} 
\usepackage{multirow} 

\def\reff@jnl#1{{\rm#1\/}}

\def\aj{\reff@jnl{AJ}}                  
\def\araa{\reff@jnl{ARA\&A}}            
\def\apj{\reff@jnl{ApJ}}                
\def\apjl{\reff@jnl{ApJ}}               
\def\apjs{\reff@jnl{ApJS}}              
\def\ao{\reff@jnl{Appl.Optics}}         
\def\apss{\reff@jnl{Ap\&SS}}            
\def\aap{\reff@jnl{A\&A}}               
\def\aapr{\reff@jnl{A\&A~Rev.}}         
\def\aaps{\reff@jnl{A\&AS}}             
\def\azh{\reff@jnl{AZh}}                        
\def\baas{\reff@jnl{BAAS}}              
\def\jrasc{\reff@jnl{JRASC}}            
\def\memras{\reff@jnl{MmRAS}}           
\def\mnras{\reff@jnl{MNRAS}}            
\def\pra{\reff@jnl{Phys.Rev.A}}         
\def\prb{\reff@jnl{Phys.Rev.B}}         
\def\prc{\reff@jnl{Phys.Rev.C}}         
\def\prd{\reff@jnl{Phys.Rev.D}}         
\def\prl{\reff@jnl{Phys.Rev.Lett}}      
\def\pasp{\reff@jnl{PASP}}              
\def\pasj{\reff@jnl{PASJ}}              
\def\qjras{\reff@jnl{QJRAS}}            
\def\skytel{\reff@jnl{S\&T}}            
\def\solphys{\reff@jnl{Solar~Phys.}}    
\def\sovast{\reff@jnl{Soviet~Ast.}}     
 \def\ssr{\reff@jnl{Space~Sci.Rev.}}     
\def\zap{\reff@jnl{ZAp}}                        
\def\nat{\reff@jnl{Nature}}             

\title[VSA Observations of the Anomalous Microwave Emission in the Perseus Region]{VSA Observations of the Anomalous Microwave Emission in the Perseus Region}

\author[C.T.~Tibbs et al.]{Christopher T. Tibbs,$\!^{1}$\thanks{E-mail:~ctibbs@jb.man.ac.uk (CTT)} Robert A. Watson,$\!^{1}$ Clive Dickinson,$\!^{1}$ Rodney D. Davies,$\!^{1}$ \and Richard J. Davis,$\!^{1}$ Carlos del Burgo,$\!^{2,3}$ Thomas M. O. Franzen,$\!^{4}$ \and Ricardo G\'{e}nova-Santos,$\!^{2,5}$ Keith Grainge,$\!^{4,6}$ Michael P. Hobson,$\!^{4}$ \and Carmen P. Padilla-Torres,$\!^{2}$ Rafael Rebolo,$\!^{2,7}$ Jos\'{e} Alberto Rubi{\~n}o-Mart{\'{\i}}n,$\!^{2,5}$ \and Richard D. E. Saunders,$\!^{4,6}$ Anna M. M. Scaife,$\!^{4}$ and Paul F. Scott$^{4}$ \\ 
$^{1}$Jodrell Bank Centre for Astrophysics, School of Physics and Astronomy, The University of Manchester, Manchester, M13 9PL, UK \\
$^{2}$Instituto de Astrofis\'{\i}ca de Canarias, 38200 La Laguna, Tenerife, Spain \\
$^{3}$UNINOVA/CA3, Campus da FCT/UNL, Qunita da Torre 2829-516 Caparica, Portugal \\
$^{4}$Astrophysics Group, Cavendish Laboratory, J.J. Thomson Avenue, Cambridge, CB3 0HE, UK \\
$^{5}$Dept. of Astrophysics, Unvi. de la Laguna, Tenerife, Spain \\
$^{6}$Kavli Institute for Cosmology, Cambridge, Madingley Road, Cambridge CB3 0HA, UK \\
$^{7}$Consejo Superior de Investigaciones Cient{\'{\i}}ficas, Spain }

\begin{document}

\date{Received **insert**; Accepted **insert**}

\pagerange{\pageref{firstpage}--\pageref{lastpage}} 
\pubyear{}

\maketitle
\label{firstpage}


\begin{abstract}
The dust feature G159.6--18.5 in the Perseus region has previously been observed with the COSMOSOMAS experiment~\citep{Watson:05} on angular scales of~$\approx$~1$^{\circ}\!$, and was found to exhibit anomalous microwave emission. We present new observations of this dust feature, performed with the Very Small Array (VSA) at 33~GHz, to help increase the understanding of the nature of this anomalous emission.

On the angular scales observed with the VSA~($\approx$~10~--~40$^{\prime}$), G159.6--18.5 consists of five distinct components, each of which have been individually analysed. All five of these components are found to exhibit an excess of emission at 33~GHz, and are found to be highly correlated with far-infrared emission. We provide evidence that each of these compact components have anomalous emission that is consistent with electric dipole emission from very small, rapidly rotating dust grains. These components contribute~$\approx$~10~\% to the flux density of the diffuse extended emission detected by COSMOSOMAS, and are found to have a similar radio emissivity.
\end{abstract}




\section{Introduction}
\label{sec:intro}

Within our own Galaxy, there are three well defined sources of diffuse Cosmic Microwave Background (CMB) foreground continuum emission, namely, synchrotron and free--free emission, which dominate at frequencies below $\approx$~60~GHz, and thermal (vibrational) dust emission, which dominates at higher frequencies. However, there is increasing evidence supporting the existence of a new continuum emission mechanism in the frequency range $\approx$~10~--~60~GHz~\citep[][and references therein]{Leitch:97, Banday:03, Dickinson:06, Davies:06, Hildebrandt:07, Dobler:08}. 

Various theories have been proposed to explain this ``anomalous'' emission, including magnetic dipole emission~\citep{DaL:99}, hard, flat spectrum, synchrotron emission~\citep{Bennett:03} and bremsstrahlung from very hot (T$_{e}$~$\sim$~10$^{6}$~K), shock heated gas~\citep{Leitch:97}. However, the currently favoured emission mechanism is that of very small (N~$\le$~10$^{3}$ atoms), rapidly rotating ($\sim$~1.5x10$^{10}$ s$^{-1}$) dust grains, emitting electric dipole radiation, commonly referred to as ``spinning dust''~\citep{DaL:98, Ali:09}. Fulleranes have also been proposed as the source of anomalous emission~\citep{Iglesias:05}.

\begin{figure*}
\begin{center}
\includegraphics*[angle=0,scale=0.7,viewport=42 90 900 515]{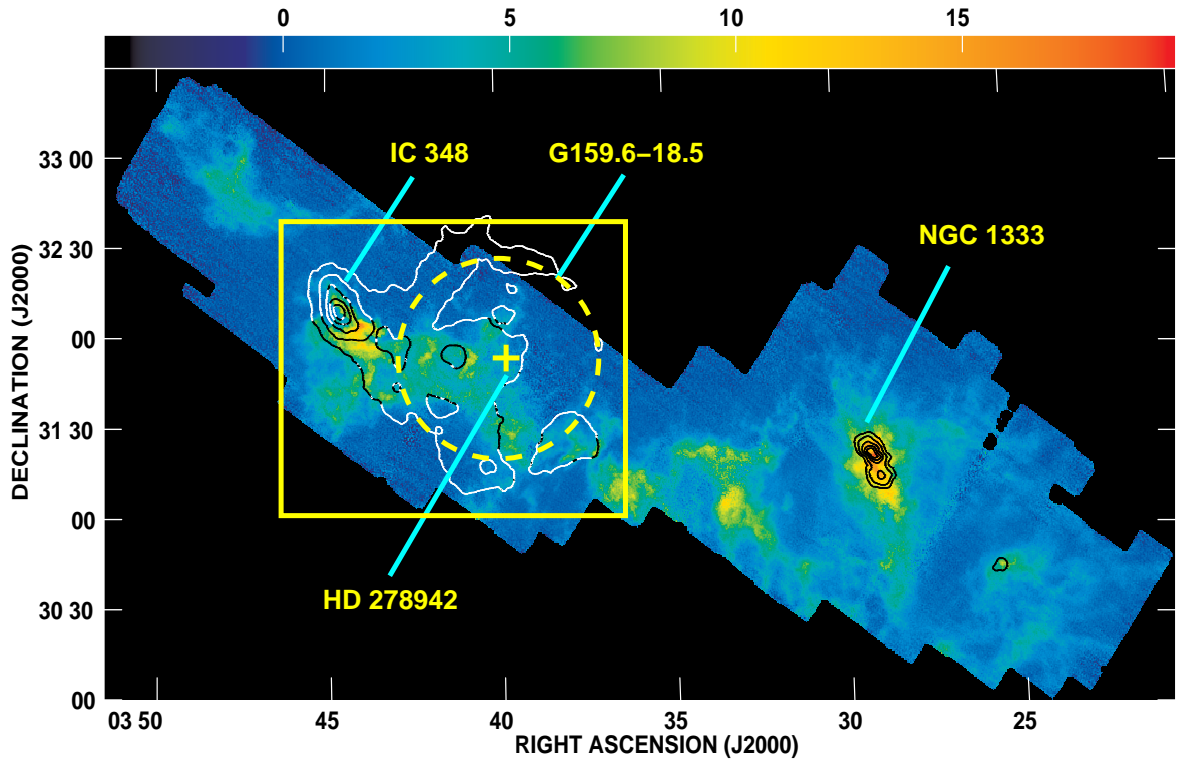}
\caption{$^{13}$CO integrated intensity false colour image~\citep{Ridge:2006} of the Perseus molecular cloud overlaid with contours of the \textit{IRIS} 100~$\mu$m emission~(10, 20, 40, 60 and 80~\% of the peak emission 680~MJy~sr$^{-1}$) showing the location of the dust shell G159.6--18.5 (dashed line), the central star HD~278942 (cross) and the location of 2 major star formation sites, IC~348 and NGC~1333. The box illustrates the region in which the VSA observations were performed.}
\label{Fig:CO}
\end{center}
\end{figure*}

This anomalous emission has been detected on large angular scales across substantial areas of the sky (\citealt{Kogut:96}; \citealt{deOC:97,deOC:99,deOC:02}; \citealt{Miville:08}) and also on small angular scales in specific Galactic objects, such as LDN~1622 (\citealt{Finkbeiner:02,Finkbeiner:04a}; \citealt{Casassus:06}; \citealt{Dickinson:07}), G159.6--18.5 (\citealt{Watson:05}), RCW~175 (\citealt{Dickinson:09}), LDN~1111 (\citealt{Ami:09}) and a number of  Lynds Dark Nebulae (\citealt{Scaife:09}) and is often found to be tightly correlated with the far-infrared (FIR) emission. The FIR emission is generally interpreted as due to three dust components: big grains (BGs), very small grains (VSGs) and polycyclic aromatic hydrocarbons (PAHs), deduced from IRAS observations~\citep{Desert:90}.

One of these Galactic objects, G159.6--18.5, is located within the Perseus molecular complex. Observations performed with the COSMOSOMAS experiment~\citep{Watson:05} identified a bright source of dust-correlated emission at~$\approx$~20~--~30~GHz, with a peaked spectrum, indicative of spinning dust. This is perhaps the best example of spinning dust emission where the peaked spectrum is well fitted by a~\citet{DaL:98} spinning dust model. These results prompted further investigations of this region, which were performed with the Very Small Array (VSA), a 14-element interferometric array operating at 33~GHz;~\citet{Watson:03} give the nominal VSA setup. In the present paper we present and discuss our results from these new observations. Section~\ref{sec:obs} describes the region under investigation and presents the VSA observations, while Section~\ref{sec:data} details the data reduction. Section~\ref{sec:ancillary} discusses the available ancillary data, which were used to help gain a better understanding of the physical conditions within the region. In Section~\ref{sec:seds} we produce spectral energy distributions (SEDs) for 5 features in the region. In Section~\ref{sec:discuss} we analyse these SEDs, investigate the correlation between the microwave and IR emission, and analyse the relationship between our VSA results and the COSMOSOMAS results~\citep{Watson:05}. Finally, our conclusions are discussed in Section~\ref{sec:con}.

\begin{table}
\centering
 \caption{Summary of the characteristics of the VSA in the super-extended array configuration.}
 \begin{tabular}{l l}
 \hline
  \textbf{Location} & Teide Observatory \\
  \textbf{Altitude} & 2400 m  \\
  \textbf{Latitude} & +28$^{o}$ 18$^{\prime}$ \\
  \textbf{Declination range} & -7$^{o}$ $<$ Dec $<$ +63$^{o}$ \\
  \textbf{No. of antennas (baselines)} & 14 (91) \\
  \textbf{No. of correlations} & 182 \\
  \textbf{T$_{sys}$ (K)} & $\approx$~35 \\  
  \textbf{Frequency (GHz)} & 33 \\
  \textbf{Bandwidth (GHz)} & 1.5 \\
  \textbf{Primary beam FWHM (arcmin)} & 72 \\
  \textbf{Synthesized beam FWHM (arcmin)} & $\approx$ 7 \\
  \textbf{Sensitivity (mJy~beam$^{-1}$ in 50~hrs)} & $\approx$~5 \\
 \hline
\end{tabular}
\label{Table:Summary_VSA}
\end{table}


\section{VSA Observations of G159.6--18.5}
\label{sec:obs}

\subsection{The G159.6--18.5 Region}
\label{sec:reg}

The Perseus molecular complex is a giant molecular cloud located in the Perseus constellation at a distance of ~$\sim$~260~pc~\citep{Cernicharo:85}.  The cloud chain is~$\sim$~30~pc in length and contains many well known sites of active star formation. The dust feature observed by~\citet{Watson:05}, G159.6--18.5, appears as a remarkably complete shell of enhanced FIR emission with diameter~$\approx$~1$^{\circ}\!$.5 as shown in~Fig.~\ref{Fig:CO}.

Fig.~\ref{Fig:CO} displays a $^{13}$CO integrated intensity image~\citep{Ridge:2006} illustrating the extent of the Perseus molecular cloud. Overlaid on the image are contours of the \textit{IRIS}\footnote{Improved Reprocessing of the \textit{IRAS} Survey~\citep{IRIS:100}.} 100~$\mu$m emission showing the location of the dust shell G159.6--18.5, which does not appear to be traced by the~$^{13}$CO integrated emission. Also overlaid is a rectangle showing the coverage of the VSA observations. 

Due to its shape, G159.6--18.5 was initially believed to be a supernova remnant~\citep{Pauls:89, Fiedler:94}, but more recent observations have determined that the source of the shell is the O9.5--B0V star, HD~278942, at its geometric centre, and that the shell is filled with H\textsc{ii} gas~\citep{Andersson:00}. Observations performed as part of the COMPLETE Survey~\citep{Ridge:06} suggested that G159.6--18.5 was indeed an expanding H\textsc{ii} bubble located on the far side of the molecular cloud. We will discuss the distribution of the ionized gas in more detail in Section~\ref{sec:low_freq}.

\subsection{VSA Observations}
\label{sec:vsa_obs}

The VSA is a CMB interferometer, situated at an altitude of 2400~m at Teide Observatory, Tenerife. It has been used to measure the CMB angular power spectrum for multipole values in the range 150~$\leq$~$\ell$~$\leq$~1500~\citep{Scott:03, Grainge:03, Dickinson:04}.

The observations of the dust feature G159.6--18.5 were performed during the period September 2005 to July 2008 at 33~GHz with the ``super-extended''  array configuration\footnote{The ``super-extended'' array configuration was the third and final array configuration for the VSA. Previously the ``compact'' and ``extended'' array configurations were in operation.}. Table~\ref{Table:Summary_VSA} provides a summary of the super-extended VSA characteristics. Observations of calibration sources have shown that the VSA interferometric pointing accuracy is better than 1$^{\prime}$. 

Due to its size ($\approx$~2$^{\circ}\!$~$\times$~1$^{\circ}$), G159.6--18.5 was observed with 11 different pointings, each with a primary beam of 1$^{\circ}\!$.2 full width at half maximum (FWHM). Each of the 11 pointings were repeated numerous times~(Table~\ref{Table:Summary_VSA_Obs}), and were then combined to provide the best possible \textit{u,v}-coverage. Table~\ref{Table:Summary_VSA_Obs} provides a summary of each of the 11 pointings, including the total number of observations, and the corresponding total observing time, for each pointing. The rms noise values in Table~\ref{Table:Summary_VSA_Obs} will be discussed in Section~\ref{sec:map}.

\begin{figure}
\begin{center}
\includegraphics*[scale=0.45, viewport=120 20 800 550]{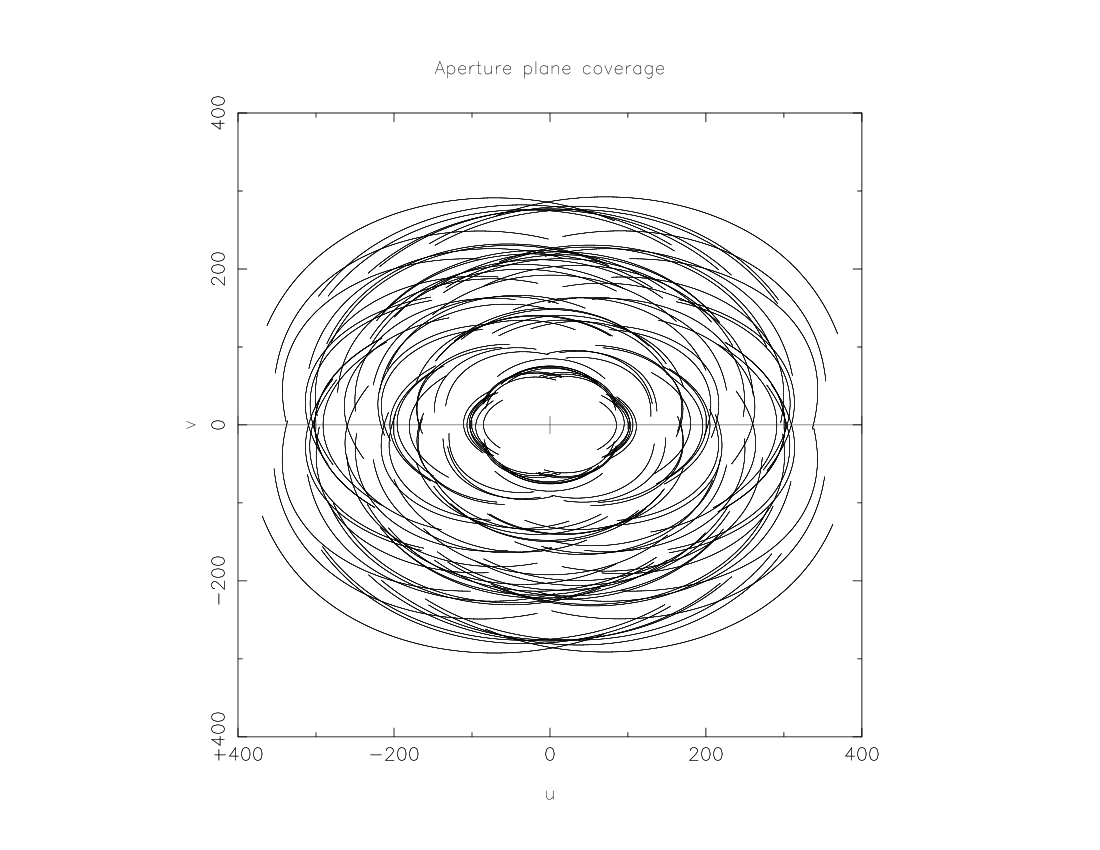}
\caption{Typical \textit{u,v}-coverage for a 6~hr VSA observation with no flagging performed. Note the central ``hole'' in the~\textit{u,v}-coverage due to the short spacing problem. X and Y axis are in units of $\lambda$.}
\label{Fig:uv}
\end{center}
\end{figure}

Fig.~\ref{Fig:uv} shows the typical \textit{u,v}-coverage for a 6~hr VSA observation with no flagging performed, and one can see the effect of the short spacing problem: the central ``hole'' in the~\textit{u,v}-coverage. A lack of short spacings occur because there is a finite limit to the smallest aperture spacing (or baseline), and this results in leaving a ``hole'' in the centre of the~\textit{u,v}-coverage. Therefore, the VSA is restricted to only observing structure on scales~$\approx$~10~--~40$^{\prime}$, and hence our observations are ``resolving out'' any larger scale emission. To alleviate this problem and observe this larger scale structure, one would need to use single dish observations with a~$\approx$~40$^{\prime}$ beam.


\section{VSA Data}
\label{sec:data}

\subsection{Data Reduction and Calibration}
\label{sec:data_red}

The VSA data reduction and calibration were performed using the software tool \textsc{reduce}. This software was specifically designed for use with VSA data and is described in detail in~\citet{Dickinson:04} and references therein. Flux density calibration was applied using Tau~A with a flux density of 350~Jy at 33~GHz. This value was determined from observations with the VSA ``extended'' array~\citep{Hafez:08} using Jupiter as the ``absolute'' calibrator with T$_{b}$~=~146.6~$\pm$~0.75~K at 33~GHz based on \textit{WMAP} observations~\citep{Hill:09}. Tau~A was chosen due to its close proximity ($\approx$~26$^{\circ}\!$ away on the sky), which meant that both the target and calibrator were available at the same time of day. Short observations of the calibrator ($\approx$~10~mins) allowed us to use Tau~A as both a flux density and phase calibrator.  Errors due to phase drifts and atmospheric/gain errors are typically less than a few percent. Throughout the paper we have assumed a conservative absolute calibration error of~5~\%. 

In addition to performing the flux density and phase calibration, \textsc{reduce} automatically performs much of the data flagging by removing antennas and baselines that were noisy or not working correctly. \textsc{reduce} also allows the user to filter out any contaminating signals such as bright sources (e.g. Sun, Moon etc.) passing through the side lobes of the VSA beam. This fringe-rate filtering is described in detail in~\citet{Watson:03}. 

After this filtering had been performed, the data were fringe-rotated to the centre of the field, and calibrated. All the data were inspected and reduced individually, and a \textit{u,v} visibility file for each observation was produced.

\begin{table*}
\begin{center}
\caption{Summary of the VSA observations.}
 \begin{tabular}{c c c c c c}
  \hline
     & & & No. of & Approx. Total & \\
  Pointing & RA & DEC & observations & Obs. Time & R.M.S. Noise \\
   & (J2000) & (J2000) & & (hr) & (mJy~beam$^{-1}$) \\
  \hline
  \hline
  A & 03:41:24 & +32:03:28 & 14 & 50 & 13.3 \\
  B & 03:46:00 & +32:03:28 & 8 & 20 & 16.4 \\
  C & 03:44:12 & +32:03:28 & 14 & 30 & 14.1 \\
  D & 03:42:48 & +32:39:50 & 19 & 65 & 16.4 \\
  E & 03:40:00 & +32:39:50 & 9 & 25 & 19.7 \\
  F & 03:38:36 & +32:03:28 & 4 & 25 & 13.9 \\
  G & 03:40:00 & +31:27:06 & 5 & 30 & 12.0 \\
  H & 03:42:48 & +31:27:06 & 7 & 25 & 14.0 \\  
  I & 03:33:32 & +31:34:33 & 14 & 45 & 19.3 \\
  J & 03:37:12 & +32:39:50 & 6 & 25 & 30.6 \\
  K & 03:37:12 & +31:27:06 & 5 & 20 & 26.8 \\  
  \hline
\end{tabular}
\label{Table:Summary_VSA_Obs}
\end{center}
\end{table*}

\subsection{VSA Map}
\label{sec:map}

\begin{figure*}
\begin{center}$
 \begin{array}{cc}
\includegraphics*[angle=0,scale=0.35,viewport= 45 10 750 580]{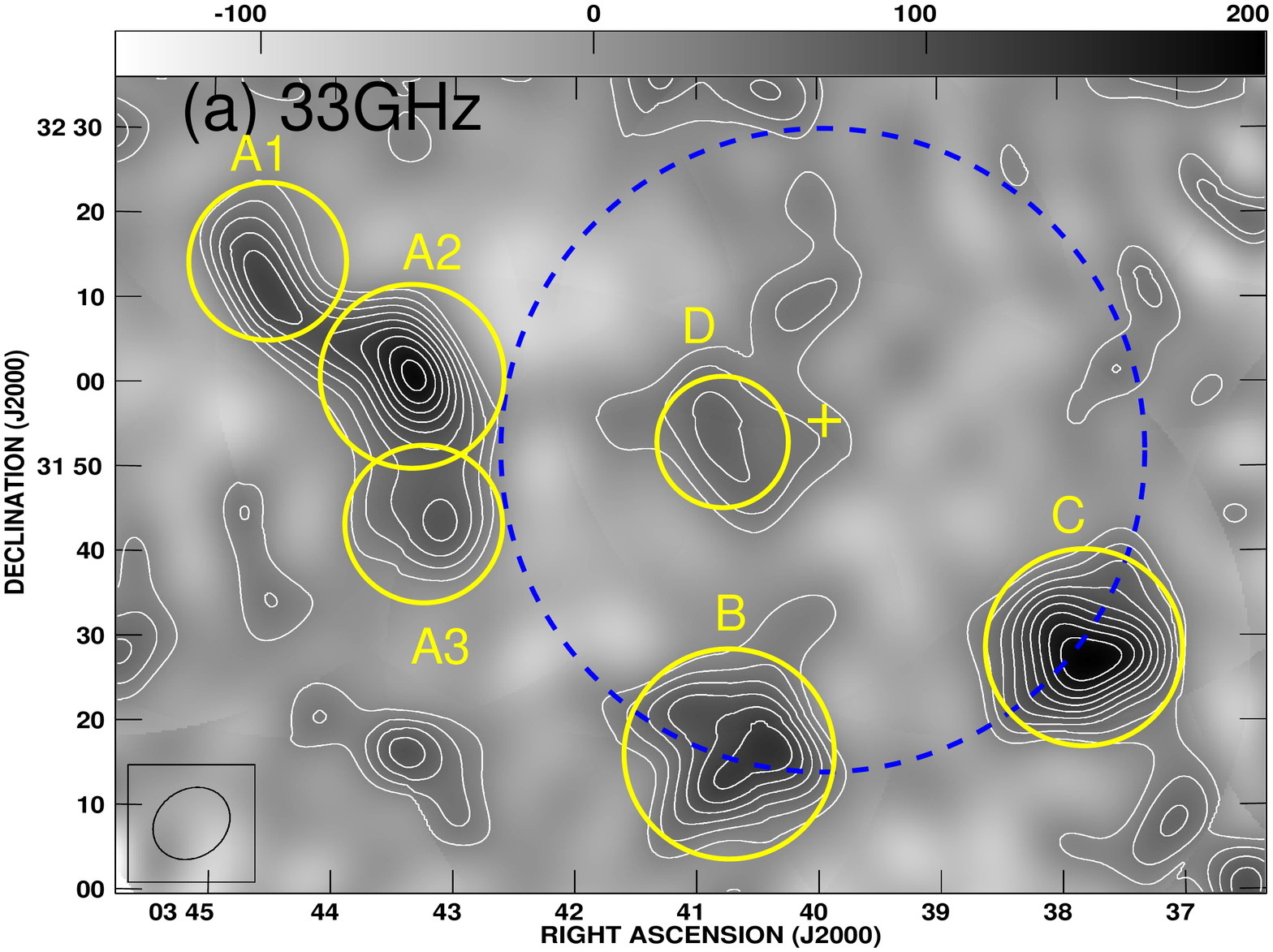} &
\includegraphics*[angle=0,scale=0.35,viewport= 45 10 750 580]{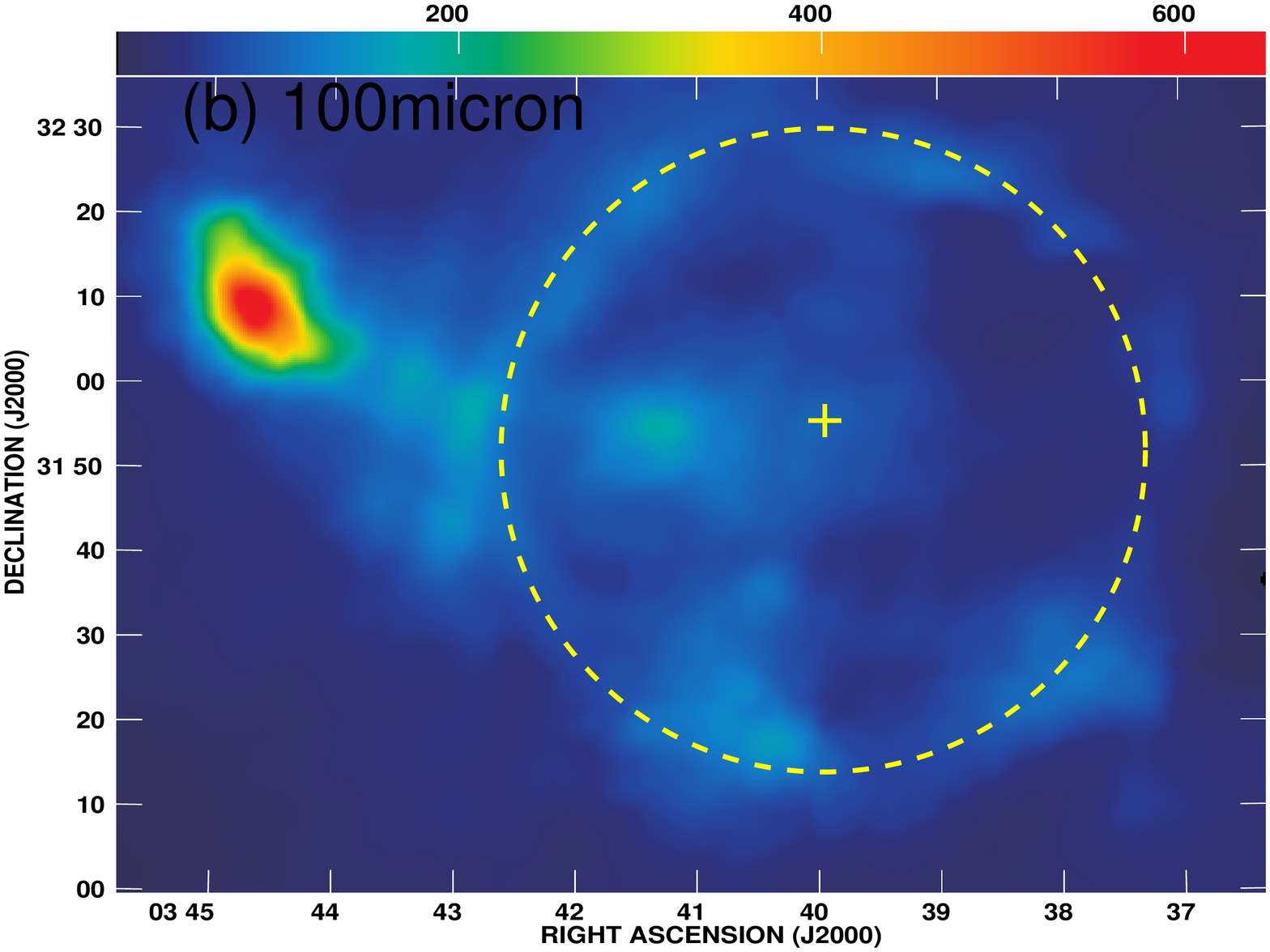} \\
  \end{array}$
  \caption{G159.6--18.5 as observed with the VSA at 33~GHz (left) and IRIS 100~$\mu$m (right). The location of the dust shell (dashed line), the O9.5--B0V star, HD~278942 (cross), and the position of the 6 features in the VSA image~(A1, A2, A3, B, C and D) are indicated. The 5 features in the VSA map, with high signal-to-noise ratios (A1, A2, A3, B and C), are discussed and analysed throughout the rest of the present paper. The central feature, D, with a much lower signal-to-noise ratio, was not included in our analysis, but will be discussed in Section~\ref{sec:high_freq}. The VSA~$\approx$~7$^{\prime}$ FWHM synthesized beam is shown in the bottom left-hand corner of the image, and the contours correspond to 10, 20, 30, 40, 50, 60, 70, 80, 90~\% of the peak flux which is 200~mJy~beam$^{-1}$. The units of the IRIS image are MJy~sr$^{-1}$.}
\label{Fig:VSA}
\end{center}
\end{figure*}

In AIPS\footnote{Astronomical Image Processing System.}, all the \textit{u,v} data for each pointing were combined to form a single \textit{u,v} visibility file for each of the 11 pointings.  A clean map was then created for each pointing using the AIPS task \textsc{imagr}, which uses a CLEAN based algorithm~\citep{Hogbom:74}. A small loop gain (0.05) was used in the cleaning process to help retain the extended emission, in which we are interested. Typical values of the rms noise in these images is~$\approx$~12~-~30~mJy~beam$^{-1}$~(Table~\ref{Table:Summary_VSA_Obs}). These rms values were calculated outside the primary beam for each image, and show that the images are not dominated by thermal noise, but are actually dominated by residual deconvolution effects. An additional contribution to the rms noise in these maps arises from the CMB. To access the levels of contamination from the CMB on the angular scales of the VSA, a simulated CMB map, based on a WMAP 5yr cosmology, was sampled with the VSA sampling distribution; this re-sampling process is described in more detail in Section~\ref{sec:GB6}. This provides a realistic level of the contribution from the CMB in these maps, and we calculated an rms of~$\approx$~5~mJy~beam$^{-1}$. This value contributes to the rms noise in the maps, but not to a significant level to provde any contaminating effects on the 5 features labelled in Fig.~\ref{Fig:VSA}a.

A mosaic of all the pointings was then created using the AIPS task \textsc{ltess}, which linearly combines each of the pointings and applies a primary beam correction in the image plane. This primary beam correction was applied out to~$\approx$~1$^{\circ}\!$~(i.e.~$\approx$~2$^{\circ}\!$ FWHM) for each individual image. The final VSA map is shown in Fig.~\ref{Fig:VSA}a with 5 features, all with a high signal-to-noise ratio, identified~(A1, A2, A3, B and C). The central feature, D, in Fig.~\ref{Fig:VSA}a has a much lower signal-to-noise ratio, and as such was not included in our analysis but it will be discussed in Section~\ref{sec:high_freq}. 

\begin{table*}
\centering
\caption{Position, integrated flux density and deconvolved angular size of the 6 features observed with the VSA at 33~GHz. Features A1, A2, A3, B and C are analysed throughout the rest of the present paper, but feature D was not included in our analysis due to the much lower signal-to-noise ratio.}
 \begin{tabular}{c c c c c c}
  \hline
  Feature & RA & DEC & Flux Density & Major Axis & Minor Axis\\
  & (J2000) & (J2000) & (Jy) & (arcmin) & (arcmin) \\
  \hline
  \hline
  A1 & 03:44:30 & +32:11:27 & 0.53~$\pm$~0.12 & 21.6~$\pm$~4.4 & 4.4~$\pm$~3.8 \\
  A2 & 03:43:27 & +32:00:17 & 1.39~$\pm$~0.20 & 24.3~$\pm$~3.1 & 12.0~$\pm$~2.2 \\
  A3 & 03:43:14 & +31:46:25 & 0.37~$\pm$~0.13 & 16.1~$\pm$~5.8 & 9.8~$\pm$~5.4 \\
  B & 03:40:41 & +31:16:31 & 1.23~$\pm$~0.21 & 20.5~$\pm$~4.8 & 17.6~$\pm$~4.8 \\
  C & 03:37:49 & +31:27:57 & 0.86~$\pm$~0.14 & 14.0~$\pm$~3.1 & 13.5~$\pm$~2.9 \\
  D & 03:40:49 & +31:54:38 & 0.18~$\pm$~0.10 & 16.0~$\pm$~9.1 & 4.9~$\pm$~5.7 \\
  \hline
\end{tabular}
\label{Table:pos}
\end{table*}

The \textit{IRIS} 100~$\mu$m image of G159.6--18.5 (Fig.~\ref{Fig:VSA}b) mostly traces the big grain dust emission in the FIR, where the dust shell is observed. One can identify that the shell consists of a few main features. Comparing Fig.~\ref{Fig:VSA}a with Fig.~\ref{Fig:VSA}b, one can see that the features observed at 33~GHz are closely correlated with the denser dust clumps in G159.6--18.5, but not with the ring structure, which does not appear in the VSA map. The spatial structure and their implications will be discussed in further detail in Section~\ref{sec:ancillary}.

The integrated flux density in each of these 6 features, shown in Fig.~\ref{Fig:VSA}a, was calculated by simultaneously fitting a gaussian component and a baseline offset using the AIPS task \textsc{jmfit}. The fitting error from \textsc{jmfit}, and the absolute calibration error of the VSA~(a conservative 5~\%~-~see Section~\ref{sec:data_red}) were added in quadrature to determine the final error. Table~\ref{Table:pos} lists the position, integrated flux density and deconvolved angular size of each of the 6 features observed with the VSA.


\section{Ancillary Data}
\label{sec:ancillary}

Due to its size and range of physical conditions, the Perseus molecular complex is a well-studied region, and therefore there are a wide variety of ancillary data available, which we used to help with our analysis~(Table~\ref{Table:ancill}). Fig.~\ref{Fig:ancill} displays the ancillary data\footnote{Data were downloaded from the \textit{Skyview} website (http://skyview.gsfc.nasa.gov), the LAMBDA website (http://lambda.gsfc.nasa.gov) and the COMPLETE Survey website (http://www.cfa.harvard.edu/COMPLETE).} overlaid with the VSA contours at 33~GHz, to help provide a picture of what different emission processes are occurring within the region.

\subsection{Radio Frequency Data}
\label{sec:low_freq}

At frequencies below $\approx$~60~GHz, the emission is expected to be dominated by synchrotron and free--free emission. However, since G159.6--18.5 is far from the Galactic plane (\textit{l}~=~159$^{\circ}\!$.6 and \textit{b}~=~$-$18$^{\circ}\!$.5) and is not in the vicinity of any known supernova remnants, the synchrotron emission is expected to be negligible at 33~GHz. Therefore the low frequency ancillary data help us to understand the free--free emission.

\begin{table}
\centering
\caption{Summary of the ancillary data used in our analysis. References are: [1]~\citet{Bonn:86}; [2]~\citet{Finkbeiner:03}; [3]:~\citet{NVSS:98}; [4]~\citet{Condon:89}; [5]~\citet{Ridge:2006}; [6]~\citet{Hinshaw:09}; [7]~ \citet{IRIS:100}.}
 \begin{tabular}{l l l c}
  \hline
 Telescope/ & Frequency/ & Angular & Ref. \\
 Survey & Wavelength & Resolution & \\
  \hline
  \hline
  Stockert & 1.4~GHz & 34$^{\prime}$ & [1] \\
  WHAM/VTSS/SHASSA & H$\alpha$ & 6$^{\prime}$ & [2] \\
  NVSS & 1.4~GHz & 0$^{\prime}\!$.75 & [3] \\
  GB6 & 4.85~GHz & 3$^{\prime}\!$.5 & [4] \\
  2MASS/NICER & Av & 5$^{\prime}$ & [5] \\
  \textit{WMAP} & 94~GHz & 12$^{\prime}\!$.6 & [6] \\
 \textit{IRAS/IRIS} & 12~$\mu$m & 3$^{\prime}\!$.8 & [7] \\
  \textit{IRAS/IRIS} & 25~$\mu$m & 3$^{\prime}\!$.8 & [7] \\
   \textit{IRAS/IRIS} & 60~$\mu$m & 4$^{\prime}\!$.0 & [7] \\
    \textit{IRAS/IRIS} & 100~$\mu$m & 4$^{\prime}\!$.3 & [7] \\
  \hline
\end{tabular}
\label{Table:ancill}
\end{table}

\begin{figure*}
\begin{center}$
 \begin{array}{cc}
\includegraphics*[scale=0.35,angle=0,viewport=45 10 750 580]{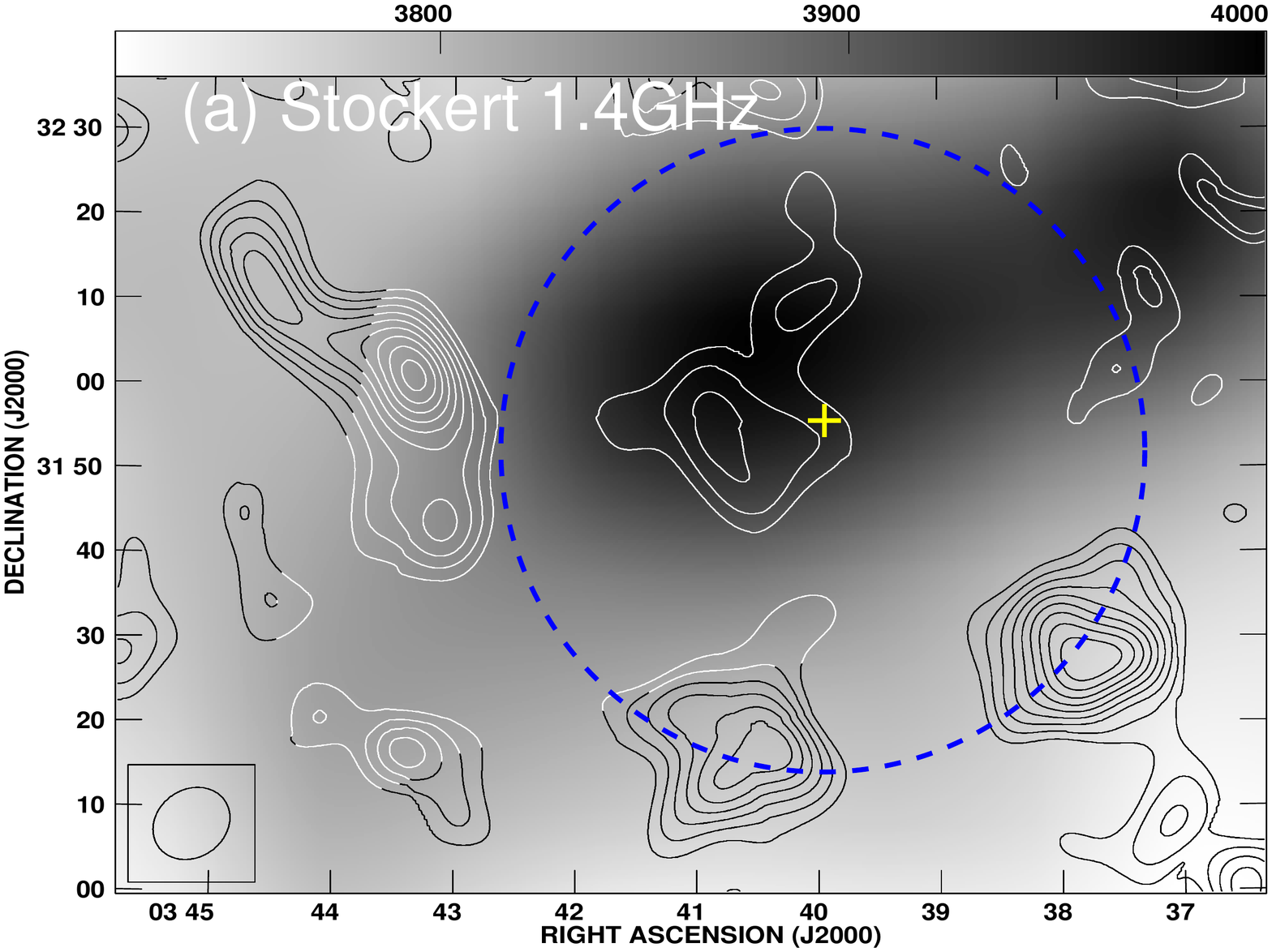} &
\includegraphics*[scale=0.35,angle=0,viewport=45 10 750 580]{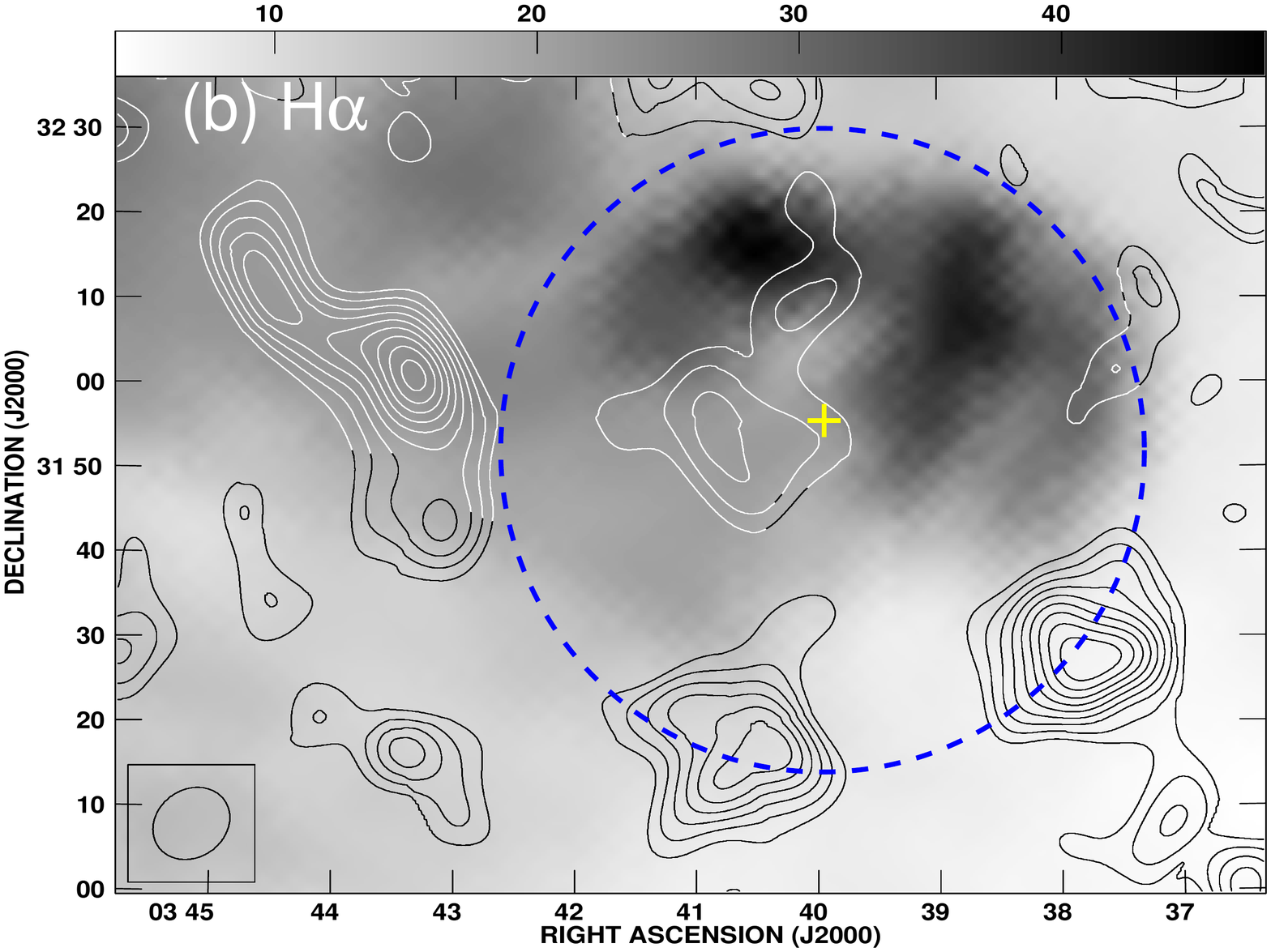} \\
\includegraphics*[scale=0.35,angle=0,viewport=45 10 750 580]{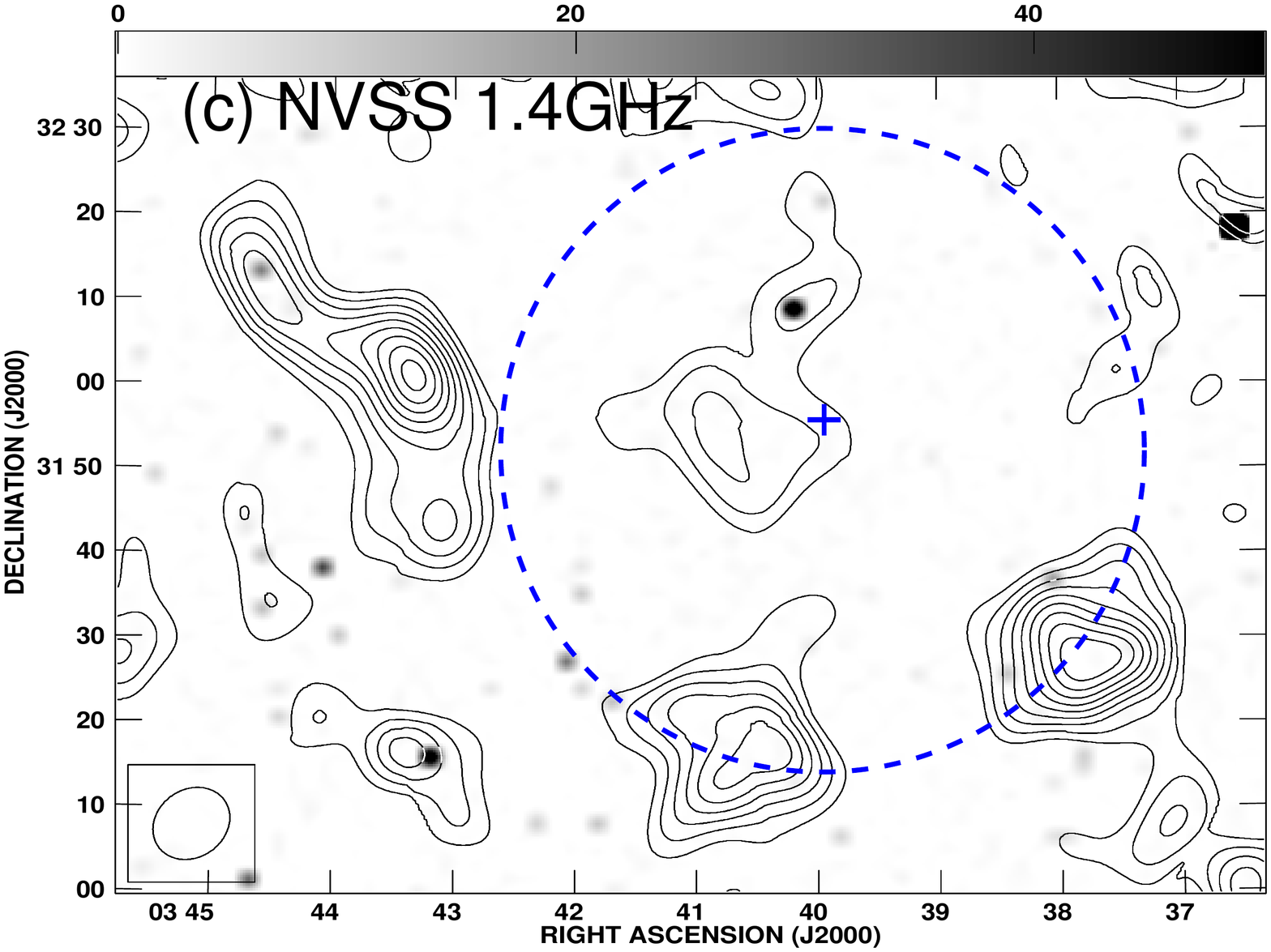} &
\includegraphics*[scale=0.35,angle=0,viewport=45 10 750 580]{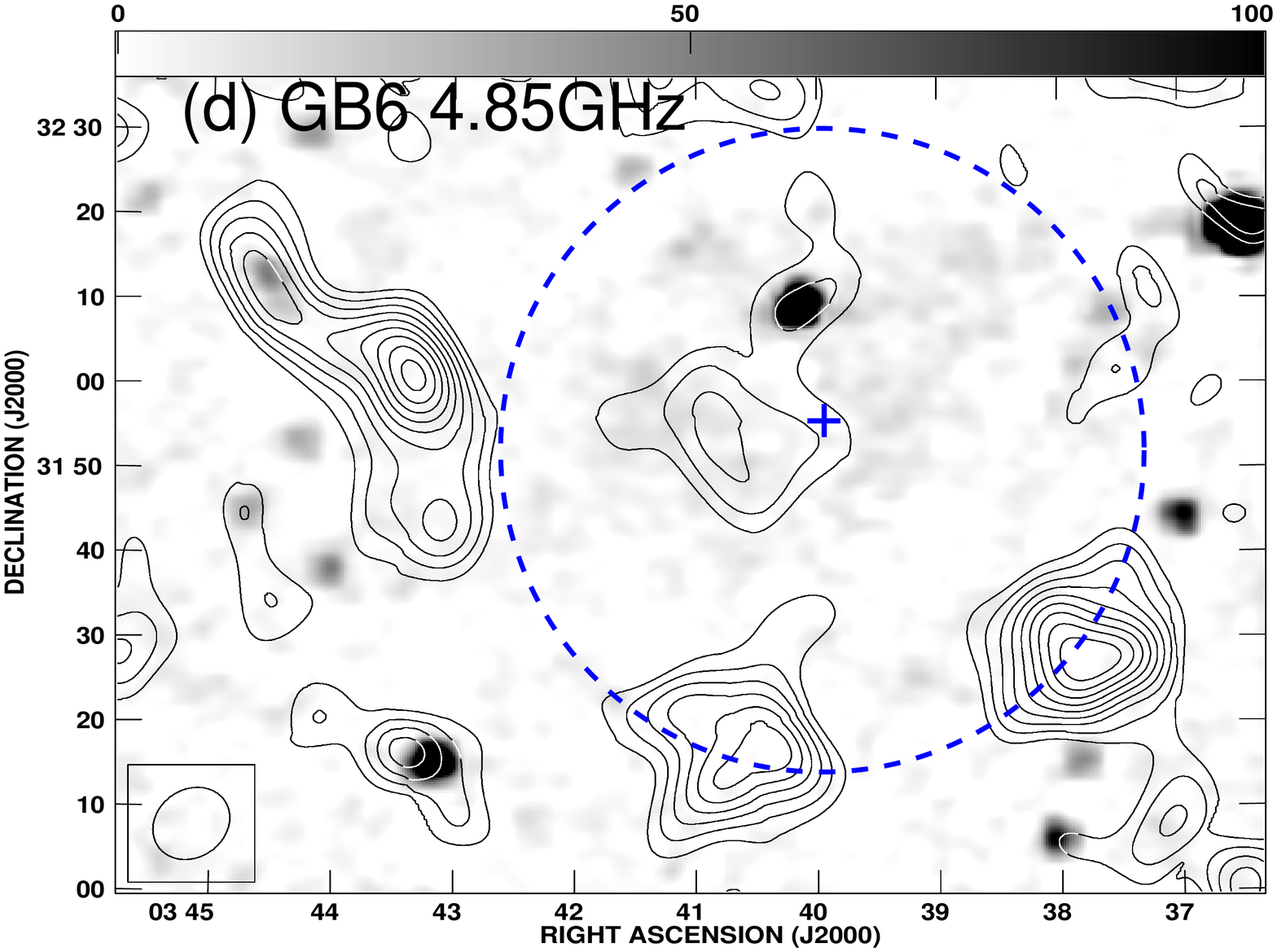} \\
\includegraphics*[scale=0.35,angle=0,viewport=45 10 750 580]{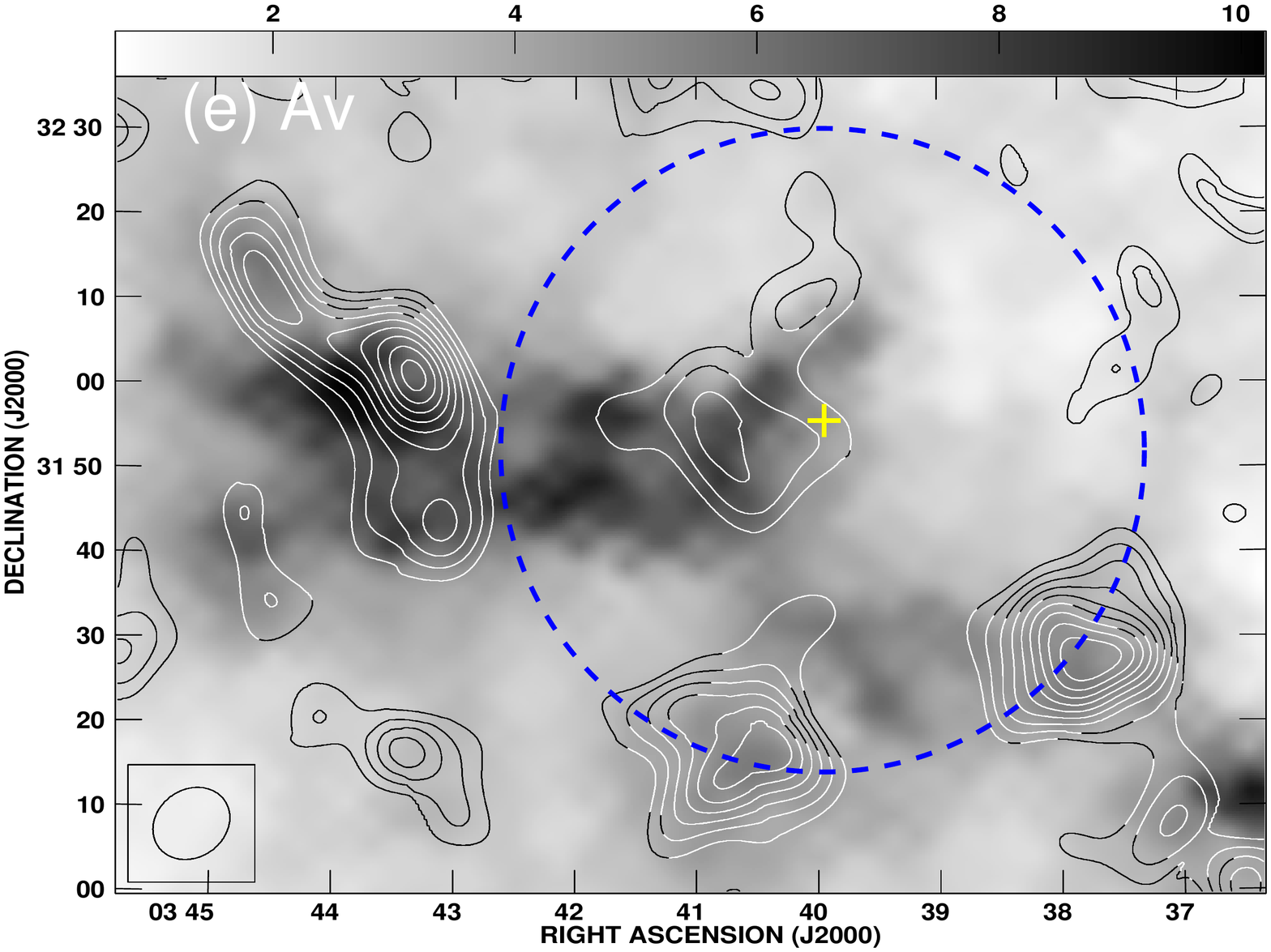} &
\includegraphics*[scale=0.35,angle=0,viewport=45 10 750 580]{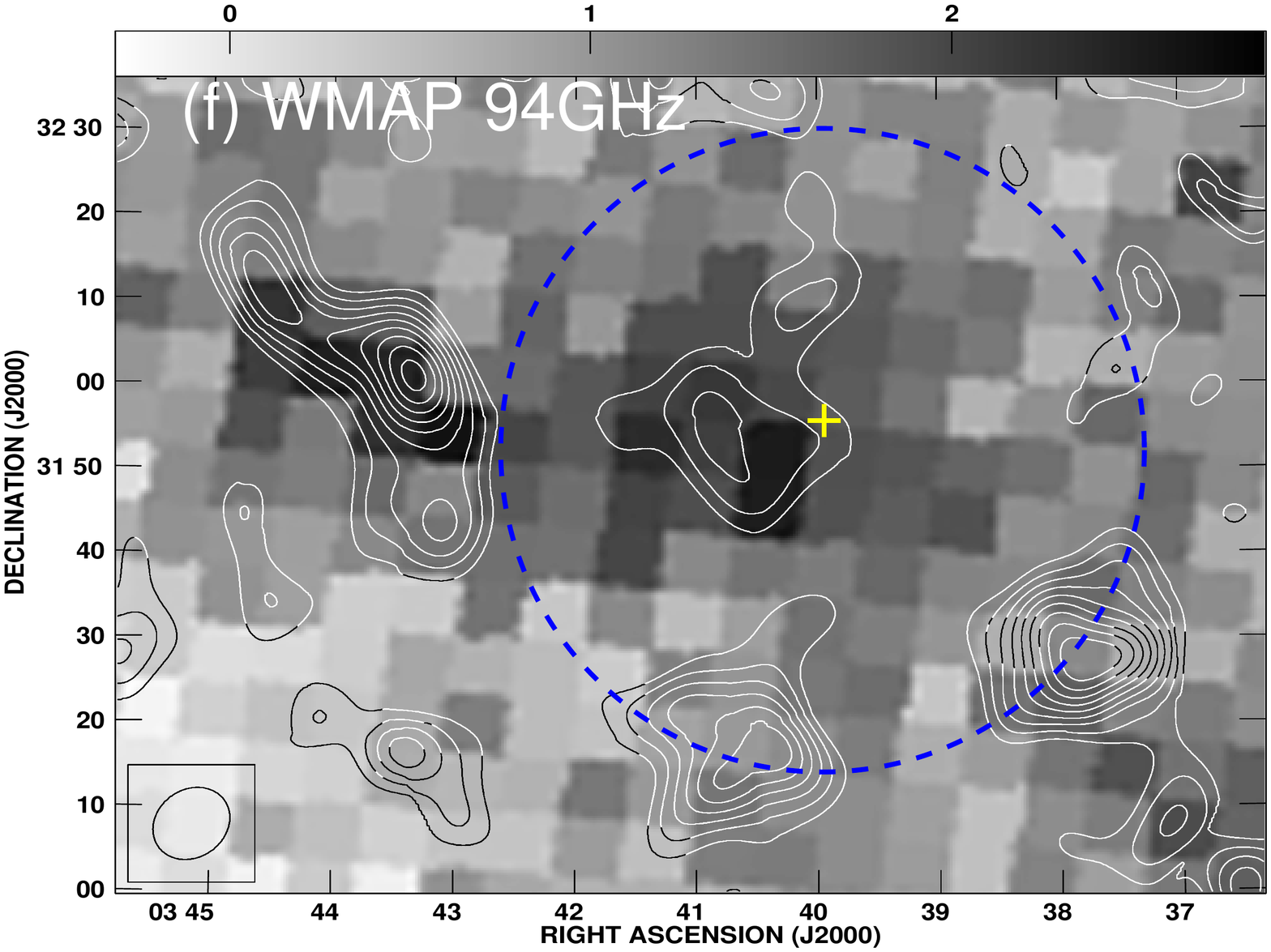} \\
  \end{array}$
  \caption{Ancillary data overlaid with the VSA contours and synthesized beam, as shown in Fig.~\ref{Fig:VSA}a, displaying the different emission mechanisms in grey scale occurring in the region: a)~Stockert 1.4~GHz in units of mK (T$_{b}$); b)~H$\alpha$ in units of R; c)~NVSS 1.4~GHz in units of mJy~beam$^{-1}$; d)~GB6 4.85~GHz in units of mJy~beam$^{-1}$; e)~Av in units of mag; f)~\textit{WMAP} 94~GHz in units of Jy~beam$^{-1}$. The dashed line displays the position of the dust shell, and the cross marks the position of the central star, HD~278942.}
     \label{Fig:ancill}
   \end{center}
\end{figure*}

In the Stockert 1.4~GHz survey~(Fig.~\ref{Fig:ancill}a), one can identify a region of diffuse emission within the shell that is offset from the centre by~$\approx$~10~-~15$^{\prime}$. This diffuse region of~$\approx$~40~arcmin diameter has also been mapped at~2.7~GHz by~\citet{ReichReich:09}. There appears to be little or no emission at 1.4~GHz from the 5 features observed at 33~GHz with the VSA. This agrees with the current understanding of the region, which is believed to contain a bubble of H\textsc{ii} gas within the dust shell~(see Section~\ref{sec:reg}). The H$\alpha$ image~(Fig.~\ref{Fig:ancill}b) appears to confirm this hypothesis, with the H$\alpha$ emission also being confined to the interior of the shell, and also offset similar to the Stockert observations, which we would expect as it traces the hot (T$_{e}$~$\sim$~10$^{4}$~K) ionized gas.

The NRAO VLA Sky Survey~(NVSS) at 1.4~GHz and the GB6 4.85~GHz survey~(Fig.~\ref{Fig:ancill}c and Fig.~\ref{Fig:ancill}d respectively) have much higher resolution than both the Stockert survey and the VSA observations, and are very useful for identifying compact radio sources, including H\textsc{ii} features. We find that there are three radio sources in the G159.6--18.5 region:~4C~31.14 to the south--east of the ring; 3C~92 within the ring; and 4C~32.14 to the north--west of the ring. Each of these sources, 4C~31.14, 3C~92 and 4C~32.14, are extra-galactic with spectral indices\footnote{Throughout this paper, the flux density spectral index is defined as~S$_{\nu}$~$\propto$~$\nu^{\alpha}\!$.}  of~$\alpha$~$\approx$~$-$0.7, $-$1.0 and $-$0.03 respectively. The flat spectrum quasar 4C~32.14 was removed from the VSA map by subtracting a CLEAN component model from the UV data. All 5 features observed with the VSA, however, appear to be free from any source contamination. Extra-galactic source counts at 15~GHz~\citep{Waldram:03} imply that the 30~GHz extra-galactic source population cannot remotely account for the 5 observed features. Also the fact that the features we observe are extended, strongly rules out the possibility of significant extra-galactic source contamination. 

Weak large scale emission, predominantly resolved out, can be seen within the shell in the GB6 image, and it is aligned with the Stockert survey observations. However, this large scale diffuse emission does not extend as far as the 5 features observed with the VSA.

\begin{figure*}
\begin{center}$
 \begin{array}{cc}
\includegraphics*[scale=0.35,angle=0,viewport=45 10 750 580]{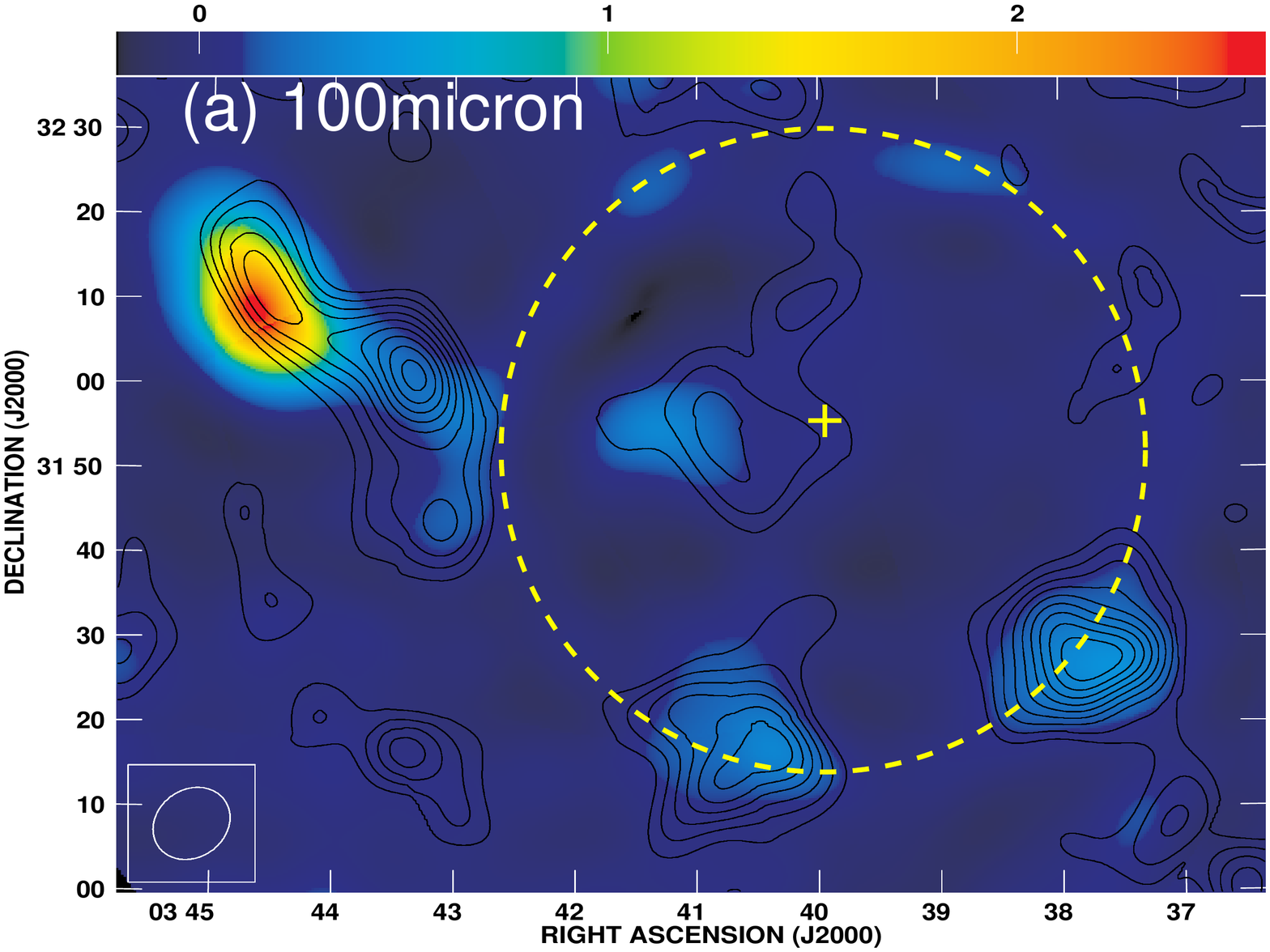} &
\includegraphics*[scale=0.35,angle=0,viewport=45 10 750 580]{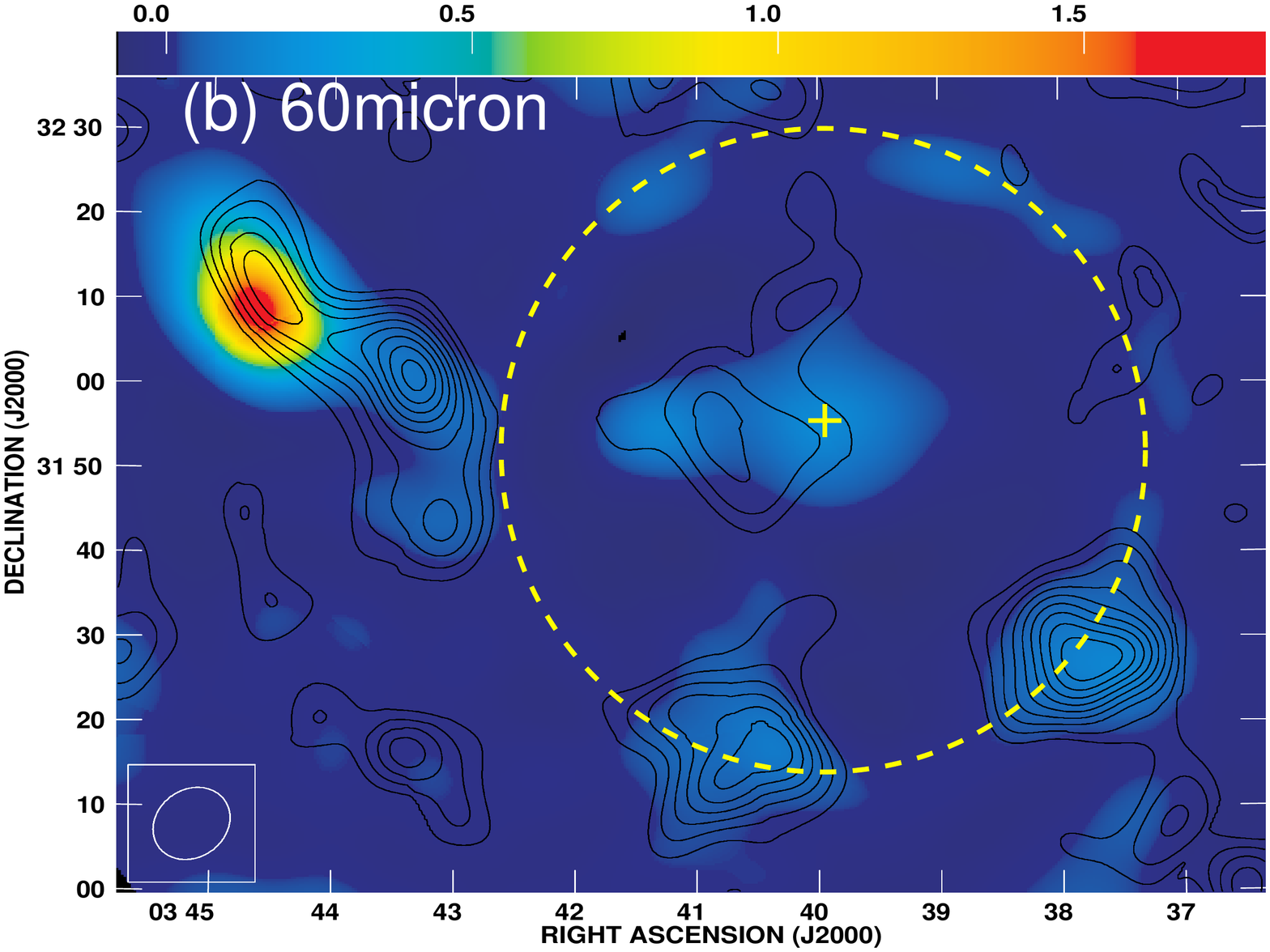} \\
\includegraphics*[scale=0.35,angle=0,viewport=45 10 750 580]{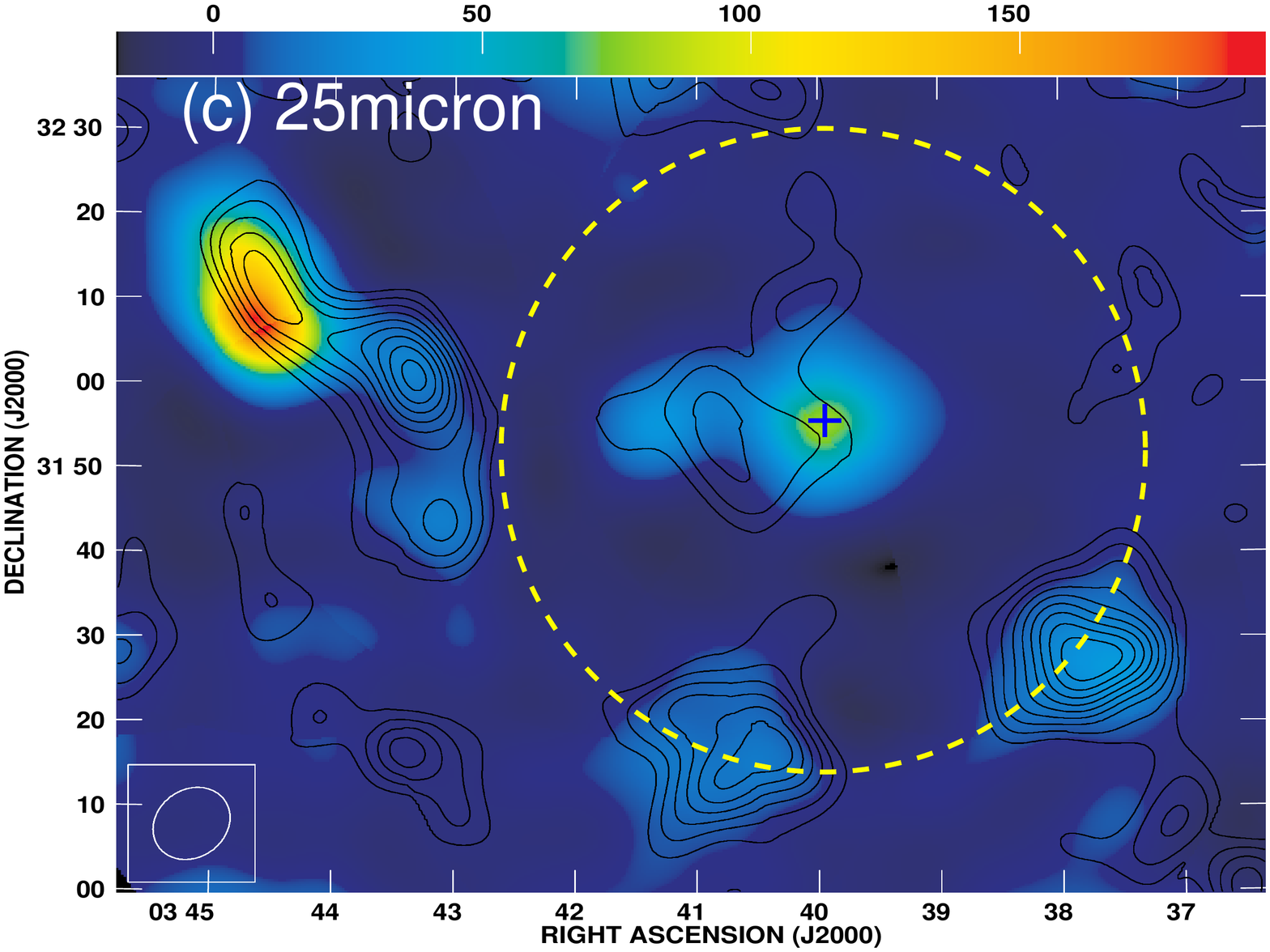} &
\includegraphics*[scale=0.35,angle=0,viewport=45 10 750 580]{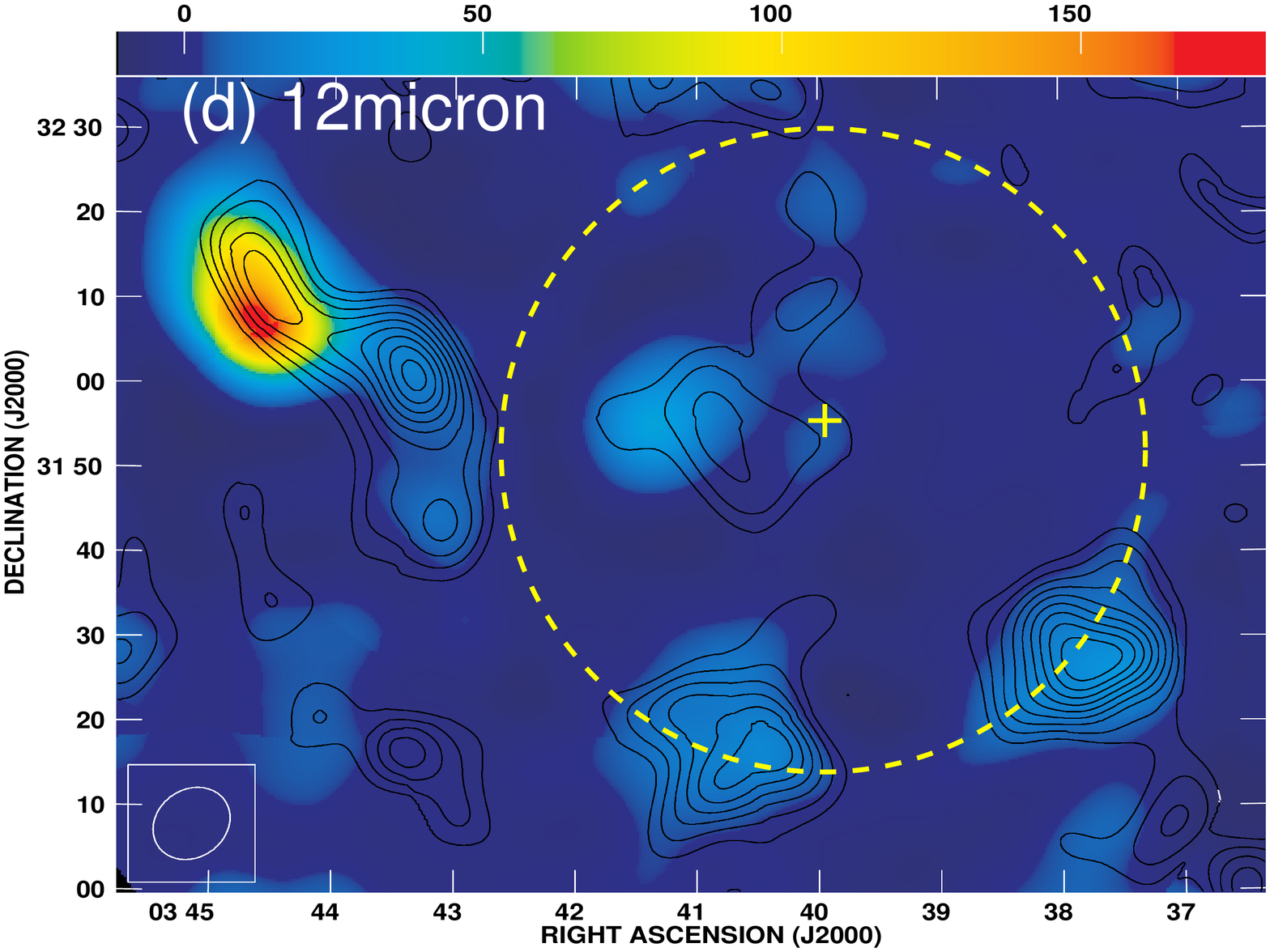} \\
  \end{array}$
  \caption{False colour \textit{IRIS} images of G159.6--18.5 processed with the VSA sampling distribution (see Section~\ref{sec:iris}) overlaid with the VSA contours and synthesized beam, as shown in Fig.~\ref{Fig:VSA}a: a) \textit{IRIS} 100~$\mu$m in units of kJy~beam$^{-1}$; b) 60~$\mu$m in units of kJy~beam$^{-1}$; c) 25~$\mu$m in units of Jy~beam$^{-1}$; d) 12~$\mu$m in units of Jy~beam$^{-1}$. The dashed line displays the position of the dust shell, and the cross marks the position of the central star, HD~278942. These images highlight the correlation between the emission at microwave and IR wavelengths. }
    \label{Fig:IRIS_image}
   \end{center}
\end{figure*}

The visual extinction (Av) map~(Fig.~\ref{Fig:ancill}e) shows that the extinction is quite high throughout the entire region ($\approx$~2~--~10~mag), suggesting a large quantity of dust in the vicinity (comparing Fig.~\ref{Fig:ancill}e with Fig.~\ref{Fig:ancill}f confirms that the extinction is due to the large, cold dust grains with T~$<$~15~K, which are traced by the 94~GHz emission). This large quantity of dust makes it very difficult to use the H$\alpha$ emission as a means to calculate an estimate of the free--free emission within the region~\citep{Dickinson:03}. The reason for this is that it is very difficult to judge what fraction of the dust, causing the extinction, is situated between us and the molecular cloud. 

However, to establish a rough estimate of whether the emission in the VSA map could be due entirely to free--free emission or not, we calculated the associated H$\alpha$ emission you would expect, assuming that all the emission observed with the VSA at 33~GHz was due to free--free emission, based on the assumptions of~\citet{Dickinson:03}. Using the conservative assumption that all the extinguishing dust is located between us and G159.6--18.5, an upper limit estimate of the actual H$\alpha$ emission was calculated. We found that the H$\alpha$ emission is much too faint~(typically on average by a factor of 20) to account for the observed 33~GHz emission. This supports the idea that the emission observed at 33~GHz is not originating entirely from free--free emission. In Section~\ref{sec:GB6} we present a more rigorous analysis that constrains the level of free--free emission observed with the VSA.

\subsection{Higher Frequency Data}
\label{sec:high_freq}

Emission at higher frequencies ($\geq$~60~GHz) is dominated by thermal dust emission, and we used the \textit{WMAP} 5-yr total intensity W-band map, and the \textit{IRIS} 100, 60, 25 and 12~$\mu$m maps to enable us to understand this emission.

The \textit{WMAP} W-band (94~GHz) image~(Fig.~\ref{Fig:ancill}f) traces the large, cold dust grains~(T~$<$~15~K) and shows that there is emission on large angular scales in the central region of G159.6--18.5. 

The IRIS images in~Fig.~\ref{Fig:IRIS_image} display the emission in the mid-IR to FIR, and highlight the high degree of correlation between the IR and the microwave emission.

The region in the centre of the shell is interesting as there appear to be two distinct FIR components, offset by~$\approx$~20$^{\prime}$. One component to the east, which is visible in all 4 \textit{IRIS} bands, and another more central component that appears to peak at~~$\approx$~60~$\mu$m, suggestive that this is much warmer~(T~$\approx$~70~K) than the surrounding dust~(T~$\approx$~30~--~40~K as displayed on the plots in Fig.~\ref{Fig:SEDs}). From Fig.~\ref{Fig:IRIS_image}c, one can see that the location of the O9.5--B0V star, HD~278942, corresponds to the brighter central component, and hence this could be acting as a heat source for the dust, and therefore producing the higher dust temperatures observed locally. 

These \textit{IRIS} images and the correlation between the microwave and the FIR emission will be discussed further in Section~\ref{sec:iris} and Section~\ref{sec:correlations} respectively.


\section{Spectral Energy Distributions of the 5 regions}
\label{sec:seds}

To identify whether the emission observed in the VSA map shown in Fig.~\ref{Fig:VSA}a is actually anomalous, we looked at the spectrum to determine whether there was an excess over the level of free--free and thermal dust emission. In order to constrain the levels of free--free emission we used the GB6 4.85~GHz observations (Section~\ref{sec:GB6}), and to constrain the levels of thermal dust emission we used the \textit{IRIS} observations in the FIR regime (Section~\ref{sec:iris}) and the \textit{WMAP} W-band data in the Raleigh--Jeans regime (Section~\ref{sec:wmap}). 

\subsection{Integrated Flux Density Spectra of Individual Features}
\label{sec:spectra}

\subsubsection{GB6}
\label{sec:GB6}

\begin{table}
\centering
\caption{Upper limits to the free--free emission in the 5 features. A 4.85~GHz 3$\sigma$ flux density upper limit based on the GB6 image. This flux density is scaled to 33~GHz using a nominal free--free spectral index of~$\alpha$~=~$-$0.12. The last column represents the percentage of free--free emission observed with the VSA at 33~GHz.}
 \begin{tabular}{c c c c c}
  \hline
   Feature & GB6 & GB6 & Filtered VSA & \% due to \\
   & S$_{4.85~GHz}$ & S$_{33~GHz}$ & S$_{33~GHz}$ & free--free \\
   & (mJy) & (mJy) & (mJy) & (3$\sigma$) \\
  \hline
  \hline
  A1 & $<$~68.3 & $<$~54.3 & 178~$\pm$~52 & $<$~31 \\
  A2 & $<$~38.8 & $<$~30.8 & 155~$\pm$~41 & $<$~20 \\
  A3 & $<$~32.9 & $<$~26.2 & 167~$\pm$~56 & $<$~16 \\
  B & $<$~41.0 & $<$~32.6 & 354~$\pm$~69 & $<$~9 \\
  C & $<$~43.1 & $<$~34.2 & 95~$\pm$~38 & $<$~36 \\
  \hline
\end{tabular}
\label{Table:GB6}
\end{table}

As discussed in Section~\ref{sec:vsa_obs}, the VSA only observes structures on angular scales of~$\approx$~10~--~40$^{\prime}$. Therefore, to determine the level of free--free emission in the VSA map, we need some low frequency observations with a similar angular resolution, namely, observations with an equivalent range of \textit{u,v}-coverage. In Section~\ref{sec:low_freq} we discussed the three available low frequency surveys: Stockert (34$^{\prime}$ FWHM at 1.4~GHz); NVSS (0$^{\prime}\!$.75 FWHM at 1.4~GHz); GB6 (3$^{\prime}\!$.5 FWHM at 4.85~GHz). None of these surveys provide an ideal match for the VSA. We cannot use the Stockert survey due to the lack of short-spacings in the VSA coverage, where all the large scale structure has been resolved out and the NVSS also suffers from an absence of short-spacings, which means that we cannot use it. This leaves the GB6 survey.

\begin{table*}
\centering
\caption{Integrated flux density for each of the 5 features identified in the VSA map at 4.85~GHz, 33~GHz, 94~GHz, 100~$\mu$m and 60~$\mu$m. All errors quoted include a fitting error and an absolute calibration error~(5~\% for VSA, 5~\% for \textit{WMAP} data and 10~\% for the \textit{IRIS} data) combined in quadrature. $^{\dag}$Flux density at 4.85~GHz is a conservative 3$\sigma$ upper limit based on the GB6 image~(see Section~\ref{sec:GB6} for details).}
 \begin{tabular}{c c c c c c}
  \hline
  Feature & S$_{4.85~GHz}$$^{\dag}$ & S$_{33~GHz}$ & S$_{94~GHz}$ & S$_{100~\mu}$$_{m}$ & S$_{60~\mu}$$_{m}$ \\
   & (Jy) & (Jy) & (Jy) & (Jy) & (Jy) \\
  \hline
  \hline
  A1 & $<$~0.20 & 0.53~$\pm$~0.12 & 0.65~$\pm$~0.63 & 5000~$\pm$~560 & 3200~$\pm$~330 \\
  A2 & $<$~0.35 & 1.39~$\pm$~0.20 & 1.99~$\pm$~0.89 & 2200~$\pm$~460 & 810~$\pm$~160 \\
  A3 & $<$~0.07 & 0.37~$\pm$~0.13 & 0.34~$\pm$~0.55 & 770~$\pm$~260 & 260~$\pm$~90 \\
  B & $<$~0.14 & 1.23~$\pm$~0.21 & 0.75~$\pm$~1.48 & 2500~$\pm$~520 & 850~$\pm$~170 \\
  C & $<$~0.39 &  0.86~$\pm$~0.14 & $<$~1.49 & 1800~$\pm$~440 & 330~$\pm$~110 \\
  \hline
\end{tabular}
\label{Table:Flux}
\end{table*}

The original GB6 observations have been filtered to remove all structure on scales larger than $\approx$~20$^{\prime}$~\citep{Condon:89} and hence cannot be used directly to assess the levels of free--free emission in the VSA image. To allow us to make use of the GB6 image, we had to filter the VSA image so as to harmonise the \textit{u,v} ranges sampled. This involved removing all the short ($\leq$ 172~$\lambda$) VSA baselines. The VSA sampling distribution was then applied to the GB6 image using the AIPS task \textsc{uvsub}. This produces a \textit{u,v} visibility file that has been sampled with the VSA sampling distribution, but with the GB6 visibilities. This was then CLEANed using the AIPS task \textsc{imagr} in a similar manner to the VSA image, creating a new GB6 image. The filtered VSA image was then directly comparable to this newly re-sampled GB6 image, and hence allowed us to determine the contribution of free--free emission on these angular scales.

The flux density of the 5 features in the filtered VSA image was calculated using \textsc{jmfit} applying both a gaussian and baseline component and is given in~Table~\ref{Table:GB6}). As discussed in Section~\ref{sec:low_freq}, all the features in the VSA image appear to show no detectable emission in the GB6 image, therefore, using the results of \textsc{jmfit} on the filtered VSA image, along with the rms noise in the same region of the re-sampled GB6 image, a 3$\sigma$~upper limit for the integrated flux density at 4.85~GHz, in each feature, was determined. This is a conservative upper limit since we are dealing with integrated fluxes rather than with peak surface brightnesses.

Table~\ref{Table:GB6} lists the integrated flux density for the 5 features at 4.85~GHz and 33~GHz, calculated as described previously in this section. The 3$\sigma$ flux density upper limit at 4.85~GHz has been scaled to 33~GHz, using a nominal free--free spectral index of $\alpha$~=~$-$0.12~\citep{Dickinson:03}. A direct comparison between the emission found in the GB6 and VSA images is given as a percentage, in the last column in Table~\ref{Table:GB6}. This suggests that the contribution of free--free emission at 33~GHz is far from dominant, and rules out the possibility that the emission in the VSA map is due entirely to free--free emission. By assuming that this ratio between the 4.85~GHz and the 33~GHz emission is independent of angular scale, this allows us to compute an estimate for the contribution of free--free emission at 33~GHz on the angular scales of the unfiltered VSA image (see Table~\ref{Table:Flux}).

\subsubsection{\textit{IRIS}}
\label{sec:iris}

The 4 \textit{IRIS} images at 100, 60, 25 and 12~$\mu$m have comparable resolution to the VSA~(Table~\ref{Table:ancill}). The VSA sampling distribution was applied to the \textit{IRIS} images using the AIPS task \textsc{uvsub}~(in a similar manner to the GB6 images as described in Section~\ref{sec:GB6}). These new \textit{IRIS} images (see Fig.~\ref{Fig:IRIS_image}) can then be directly compared with the VSA image. \textsc{jmfit} was performed on these images to calculate the integrated flux density in each of the 5 features in the VSA map. This was performed using the same \textsc{jmfit} parameters obtained from the VSA image, and the fitting error and the absolute calibration error (conservatively assumed to be~$\approx$~10~\%) were combined in quadrature to determine the overall error~(see Table~\ref{Table:Flux}).

\subsubsection{\textit{WMAP} W-band}
\label{sec:wmap}

The 5-yr \textit{WMAP} W-band map was used to constrain the Raleigh--Jeans tail of the thermal dust emission. As discussed in Section~\ref{sec:high_freq} most of the emission at 94~GHz occurs within the central region of G159.6--18.5, and extends to IC~348. To calculate the integrated flux density of each of the 5 features at 94~GHz, the VSA image was smoothed to a 12$^{\prime}\!$.6 resolution, and then using \textsc{jmfit} each of the features was fitted with a gaussian component. The size and position of these gaussians were then kept fixed, and fitted for in the \textit{WMAP} W-band image. This accounted for the difference in angular resolution between the two images.

\subsection{SED Determination}

\begin{figure*}
\begin{center}$
 \begin{array}{cc}
 \includegraphics*[scale=0.35,angle=180,viewport=50 40 750 550]{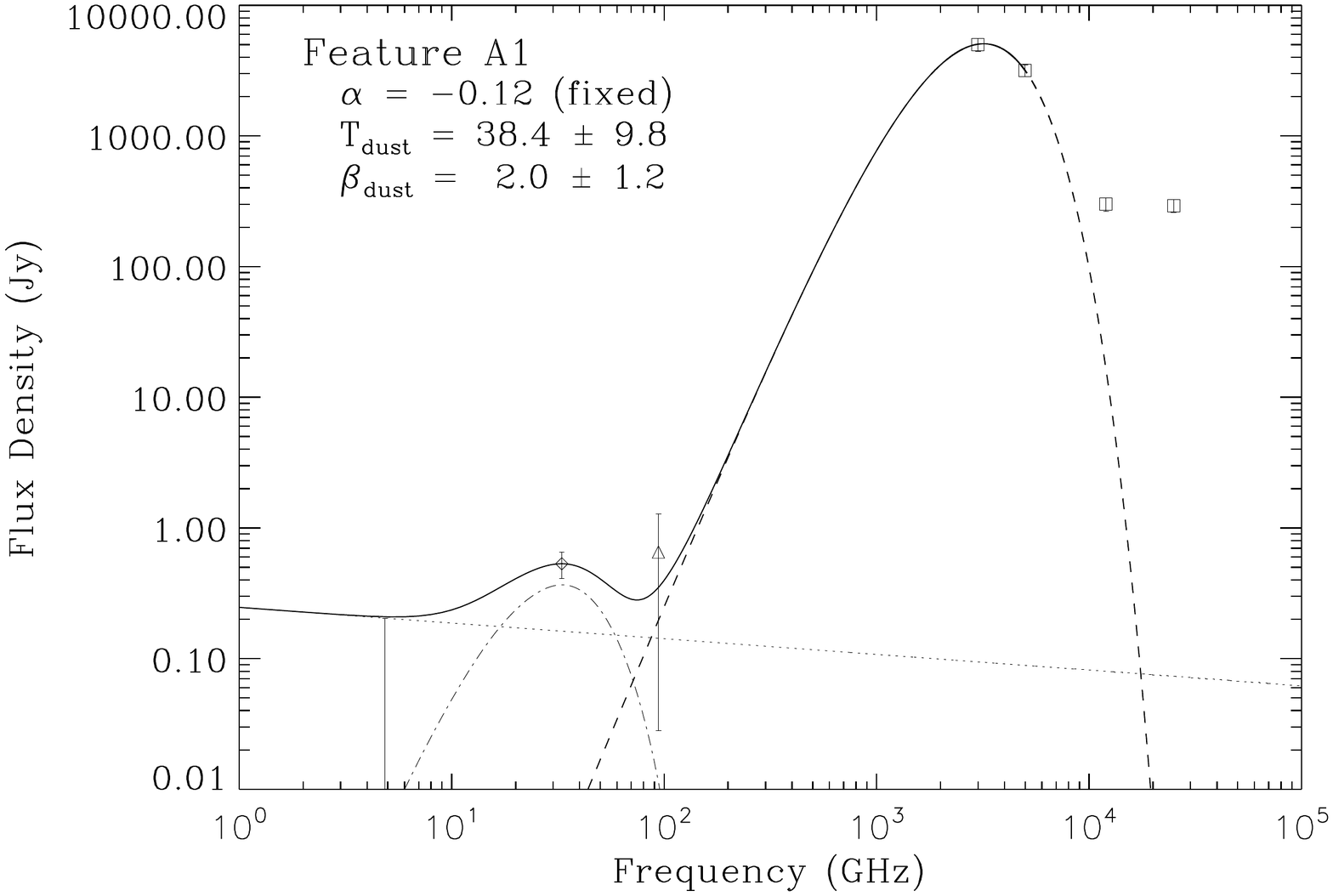} &
 \includegraphics*[scale=0.35,angle=180,viewport=50 40 750 550]{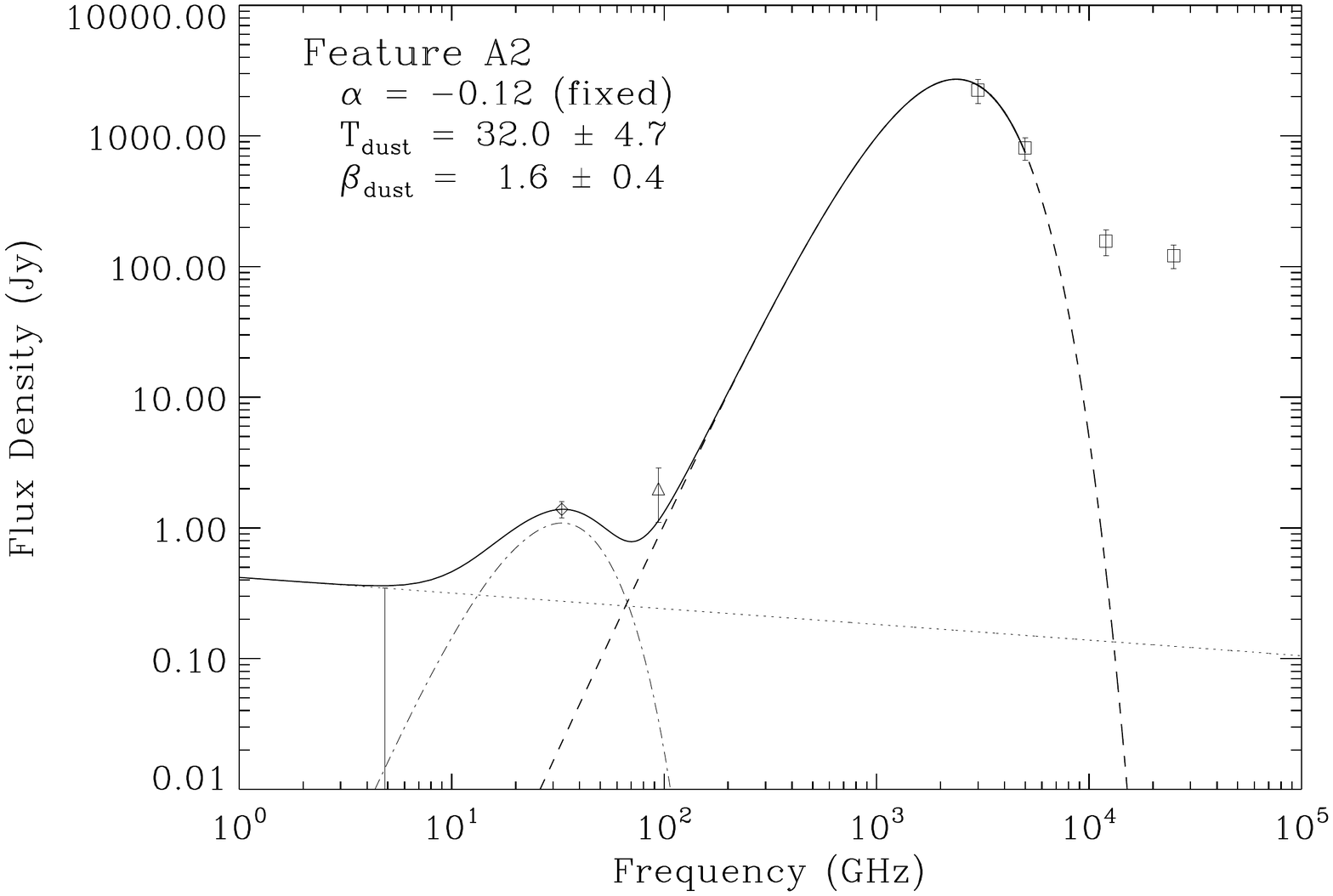} \\
 \includegraphics*[scale=0.35,angle=180,viewport=50 40 750 550]{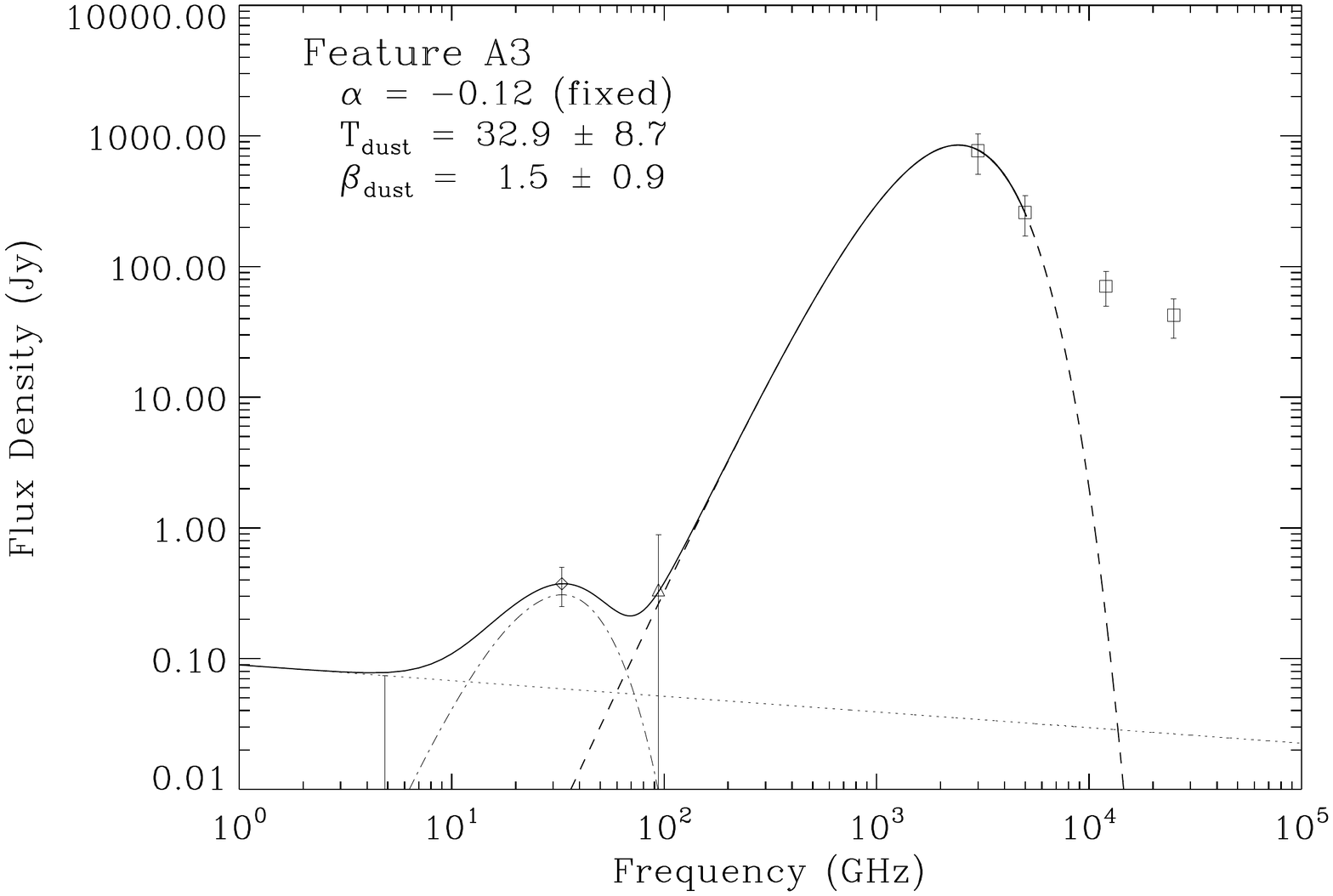} &
 \includegraphics*[scale=0.35,angle=180,viewport=50 40 750 550]{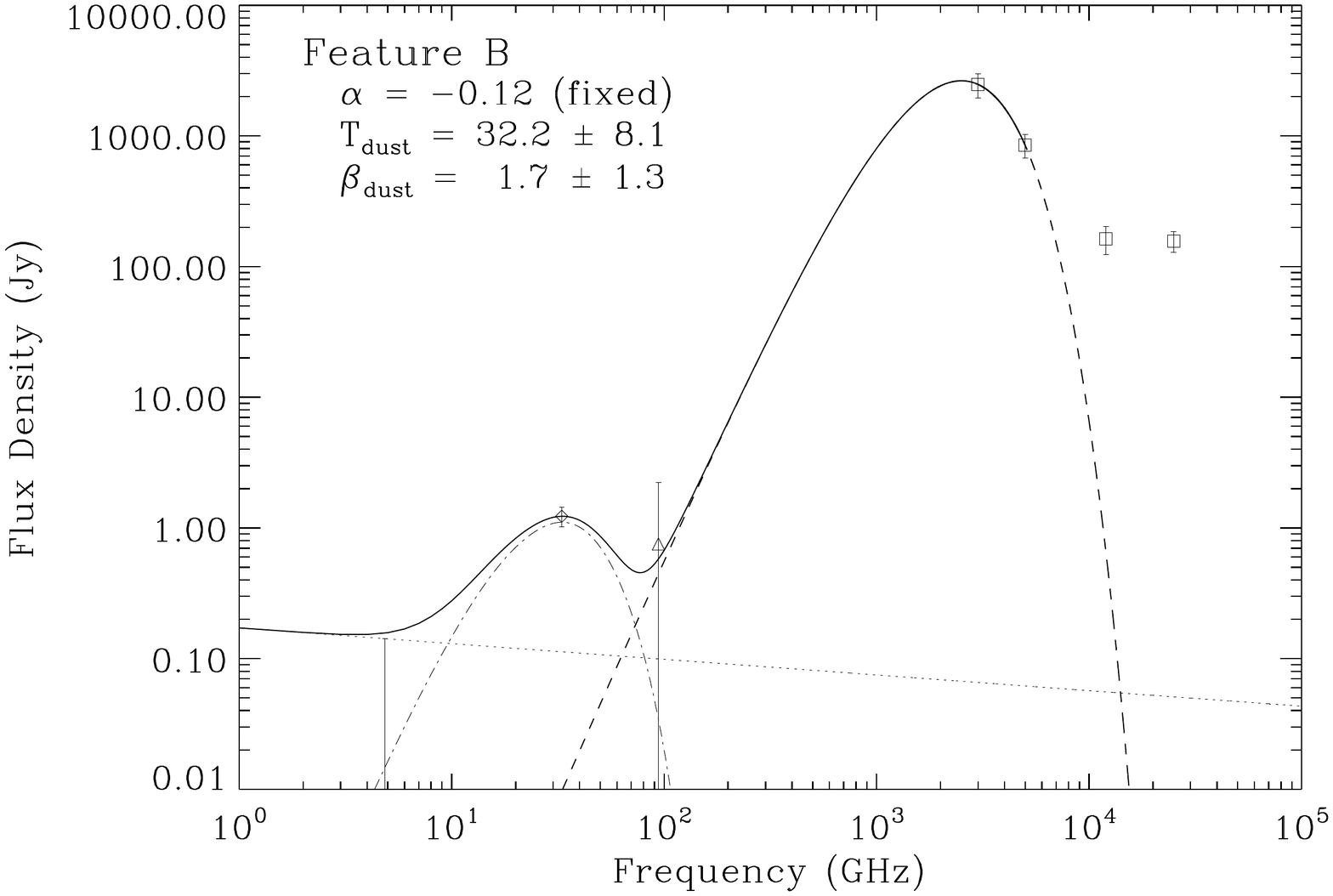} \\
 \end{array}$
 \includegraphics*[scale=0.35,angle=180,viewport=50 40 750 550]{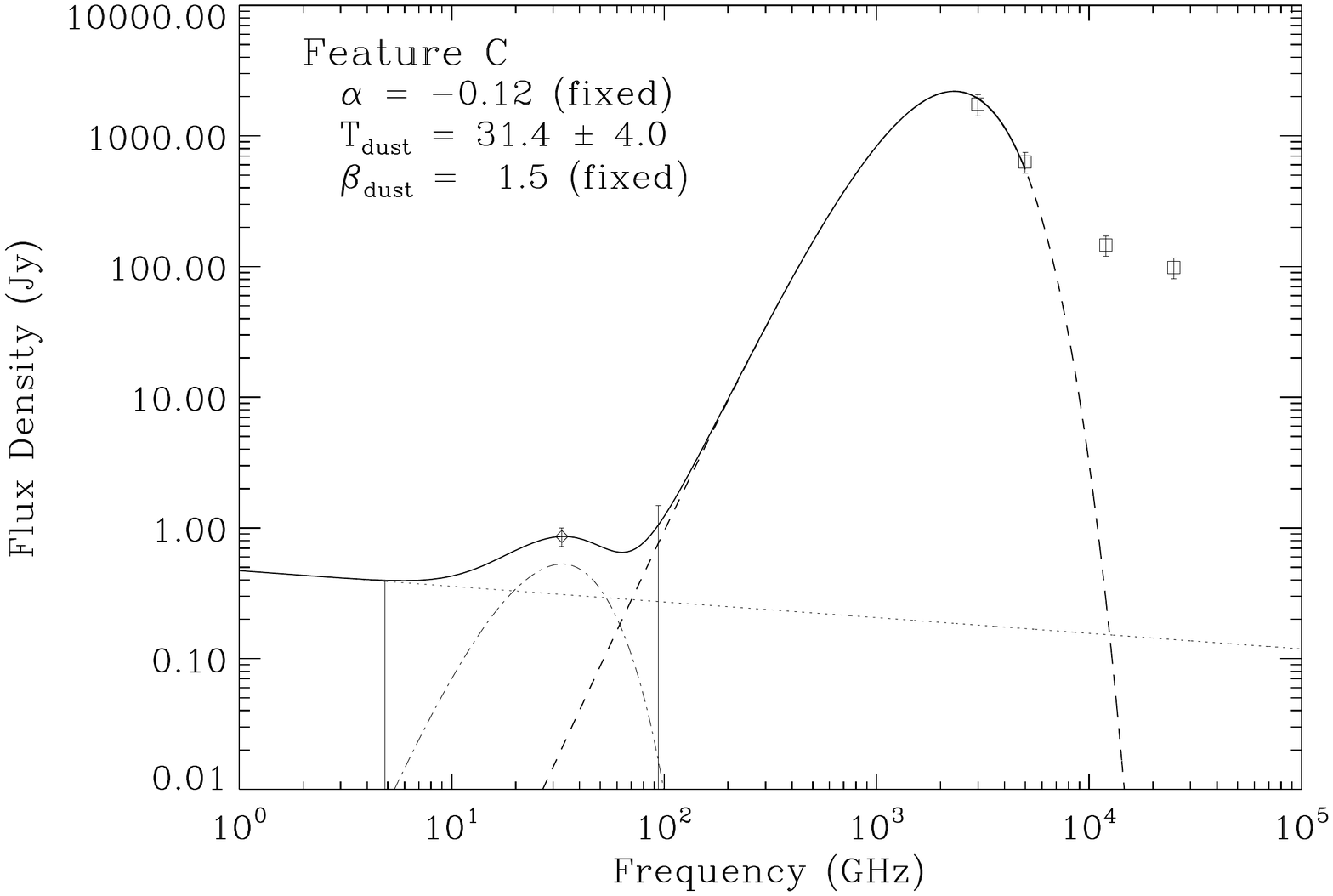} \\
  \end{center}
 \centering
  \caption{Integrated flux density spectra for the 5 features shown in Fig.~\ref{Fig:VSA}a. A power-law with spectral index~$\alpha$~=~$-$0.12 is plotted to represent the free--free emission 3$\sigma$ upper limit (dotted line) and a modified black-body curve, fitted to the 94~GHz, 100~$\mu$m and 60~$\mu$m data points, is plotted to represent the thermal dust emission (dashed line). The best fitting values for T$_{dust}$ and~$\beta_{dust}$ are displayed on each of the plots. For feature C, the 94~GHz data point was only an upper limit, and therefore the modified black-body curve is fitted with a fixed dust emissivity value of~$\beta_{dust}$~=~1.5. The VSA 33~GHz data point is not accounted for by either of these curves and is fitted by a linear combination~(0.5CNM~$+$~0.5MC) of~\citet{DaL:98} spinning dust models~(dash-dot line). Also plotted are the 25~$\mu$m and 12~$\mu$m data points, but these were not included in the SED fitting procedure.}
  \label{Fig:SEDs}
\end{figure*}

Fig.~\ref{Fig:SEDs} shows the flux density spectrum for each of the 5 features identified in Fig.~\ref{Fig:VSA}a. In Section~\ref{sec:ancillary} we discussed the emission mechanisms expected to occur within this region using the ancillary data available. Free--free emission occurs at the radio frequencies, and thermal (vibrational) dust is expected to occur at FIR wavelengths. 

We determined an upper limit for the free--free emission at 4.85~GHz, and since we know that a typical free--free emission spectrum follows a power-law with spectral index~$\alpha$~$=$~$-$0.12, we plotted this normalized to our upper limit at 4.85~GHz. We note that at frequencies below~$\approx$~1~GHz, the free--free emission can become optically thick ($\tau$~$>$~1) with a spectral index of~$\alpha$~$\approx$~$+$2, but at higher frequencies, the free--free emission becomes optically thin ($\tau$~$\ll$~1) and follows the standard power-law with spectral index~$\alpha$~$=$~$-$0.12.

For the thermal dust emission, we used a modified black-body curve function, $\nu^{\beta_{dust}+2}$B($\nu$,T$_{dust}$), where~$\beta_{dust}$ is the dust emissivity index and B($\nu$,T$_{dust}$) is the black-body function for a dust temperature, T$_{dust}$. A single temperature dust component, modelled by this modified black-body curve, was fitted to the 94~GHz, 100~$\mu$m and 60~$\mu$m data points for each feature individually. The best fit values for both~$\beta_{dust}$ and T$_{dust}$ are displayed on each of the plots in Fig.~\ref{Fig:SEDs}. 

Also plotted in Fig.~\ref{Fig:SEDs} are the 25 and 12~$\mu$m data points. These were not fitted by the single temperature modified black-body curve because the emission at these wavelengths is not due to the same dust grain population as the 100 and 60~$\mu$m. The 100 and 60~$\mu$m emission is dominated by BGs, while the 25 and 12~$\mu$m emission is dominated by VSGs. The 12~$\mu$m emission has an additional contribution from PAHs.


\section{Discussion}
\label{sec:discuss}

\subsection{SED Modelling}
\label{sec:model}

In each plot in Fig.~\ref{Fig:SEDs}, it is possible to identify an excess of emission present at 33~GHz that cannot be explained by either free--free or thermal dust emission. Table~\ref{Table:Excess} displays this fraction of excess emission at 33~GHz, and the corresponding significance, for each of the five features, confirming that a substantial fraction of the emission at 33~GHz is anomalous.

For features A1, A2, A3 and B, the 94~GHz data point provides a constraint on the shape of the thermal dust curve, providing the best fitting values for the dust  temperature~T$_{dust}$, and the dust emissivity index, $\beta_{dust}$ (see plots in Fig.~\ref{Fig:SEDs}). However, for feature C, the 94~GHz data point is an upper limit and hence does not provide a reliable constraint on the thermal dust curve. Therefore, for this feature, we used a fixed value of~$\beta_{dust}$~=~1.5. This is a typical value of~$\beta_{dust}$~\citep{Dupac:03}, and to ensure that selecting this value was reasonable, we investigated using a value of~$\beta_{dust}$~=~1.1~(towards the lowest measured value for~$\beta_{dust}$~\citep{Dupac:03}), and found that this only reduced the fraction of excess emission at 33~GHz from 62~\% to 48~\%. This confirms that there is an excess of emission observed with the VSA at 33~GHz that cannot be explained by either free--free or thermal dust emission. Therefore to fit the 33~GHz data point, a peaked spectrum is required.

\begin{table}
\centering
\caption{Fraction of excess emission at 33~GHz, and the significance, in each of the five features.}
 \begin{tabular}{c c c}
  \hline
  Feature & Fraction of & Significance \\
   & excess emission & \\
  \hline
  \hline
  A1 & $>$~0.69 & $>$~3.0$\sigma$ \\
  A2 & $>$~0.79 & $>$~5.6$\sigma$ \\
  A3 & $>$~0.82 & $>$~2.5$\sigma$ \\
  B & $>$~0.90 & $>$~5.2$\sigma$ \\
  C & $>$~0.62 & $>$~3.8$\sigma$ \\
  \hline
\end{tabular}
\label{Table:Excess}
\end{table}

Peaked spectra can be caused by emission from ultra-compact (UC) H\textsc{ii} regions, gigahertz-peaked spectra (GPS), magnetic dipole emission or electric dipole emission. UCH\textsc{ii} regions are very compact, optically thick, H\textsc{ii} regions while GPS are radio sources at high redshifts with synchrotron self--absorption. A detailed search of the \textit{IRAS} point source catalogue~\citep{IRAS:88} revealed that there are only a few sources~($<$~4~\%), within 1$^{\circ}\!$ of the centre of our image, that meet the colour criterion set by~\citet{Wood:89a} for UCH\textsc{ii} regions. Further analysis reveals that to fit the data with emission from an UCH\textsc{ii} region would require an emission measure of~$\approx$~2~$\times$~10$^{9}$~cm$^{-6}$pc and an angular size of~$\leq$~1$^{\prime\prime}$. These parameters are towards the upper end of the observed values for UCH\textsc{ii} regions~\citep{Wood:89b}, and also imply that UCH\textsc{ii} regions would appear as point--like sources in our image, which is not the case.We would also expect GPS to appear as point--like sources in our image, and to be visible in the NVSS and GB6 surveys discussed in Section~\ref{sec:low_freq}. Neither of these are true. 

Magnetic dipole emission has been ruled out due to polarization observations of the region by~\citet{Battistelli:06}. 

Therefore, having ruled out UCH\textsc{ii} regions, GPS and magnetic dipole emission as viable options, we have fitted a~\citet{DaL:98} spinning dust model for electric dipole radiation. Since we only have the one data point at 33~GHz, we fitted a linear combination of the models for the cold neutral medium and the dense molecular cloud~(0.5CNM~$+$~0.5MC) to ensure the peak occurred at 33~GHz. Clearly more data in the frequency range 5~--~60~GHz are required to confirm this fit.

\subsection{Dust Emissivities}
\label{sec:emiss}

The dust emissivity, the ratio between the microwave and IR emission, provides a convenient way\footnote{The dust emissivity is calculated relative to the 100~$\mu$m emission, and is therefore dependent on the temperature of the dust. } to represent the excess emission found in Fig.~\ref{Fig:SEDs} and is measured in units of~$\mu$K~(MJy~sr$^{-1}$)$^{-1}$. Using the values in Table~\ref{Table:Flux}, the dust emissivity between 33~GHz and 100~$\mu$m was calculated for each of the 5 features and are tabulated in Table~\ref{Table:emiss}. Also listed in Table~\ref{Table:emiss} is the dust emissivity for 6 other sources: 2 values based on H\textsc{ii} regions and 4 values for typical high Galactic latitudes.

\begin{table}
\centering
\caption{Dust emissivities at~$\approx$~30~GHz relative to 100~$\mu$m for the 5 features observed in this work, and 6 other sources to provide a comparison. The 6 H\textsc{ii} regions is a mean value based on 6 southern H\textsc{ii} regions, LPH~96 is also an H\textsc{ii} region and LDN~1622 is a dark cloud. The all-sky value is a mean value outside the Kp2 mask, and the 15 regions value is a mean value of 15 high latitude regions both from \textit{WMAP}. G159.6--18.5 is the value obtained with the COSMOSOMAS observations.}
 \begin{tabular}{l l l}
  \hline
  Source & Dust Emissivity & Reference \\
   & $\mu$K (MJy sr$^{-1}$)$^{-1}$ &  \\
  \hline
  \hline
  A1 & 2.8~$\pm$~0.7 & This paper \\
  A2 & 16.4~$\pm$~4.1 & This paper \\
  A3 & 12.8~$\pm$~6.1 & This paper \\
  B & 13.2~$\pm$~3.6 & This paper \\
  C & 13.0~$\pm$~3.2 & This paper \\
  6 H\textsc{ii} regions & 3.3~$\pm$~1.7 &\citet{Dickinson:07} \\
  LPH96 & 5.8~$\pm$~2.3 &\citet{Dickinson:06} \\
  All--sky & 10.9~$\pm$~1.1 &\citet{Davies:06} \\
  15 regions & 11.2~$\pm$~1.5 &\citet{Davies:06} \\
  LDN1622 & 24.1~$\pm$~0.7 &\citet{Casassus:06} \\
  G159.6--18.5 & 15.7~$\pm$~0.3 &\citet{Watson:05} \\
  \hline
\end{tabular}
\label{Table:emiss}
\end{table}

The radio emissivity of dust at~$\approx$~30~GHz is known to have a typical value of~$\approx$~10~$\mu$K~(MJy~sr$^{-1}$)$^{-1}$ with a factor of~$\approx$~2 variation at high Galactic latitudes~\citep{Davies:06}. The dust emissivity for the 5 features analysed in this work are consistent with this value. Comparing the dust emissivities listed in Table~\ref{Table:emiss}, we find that the features A2, A3, B and C have a dust emissivity similar to the typical value found for high Galactic latitudes, while feature A1, with a much lower dust emissivity, is similar to that for H\textsc{ii} regions. 

The lower dust emissivity per unit of 100~$\mu$m surface brightness in H\textsc{ii} regions (see Table~\ref{Table:emiss}) may be the result of the relative difference in physical environmental conditions compared with those in the general ISM. For example, the warmer dust resulting from the strong UV radiation field in an H\textsc{ii} region will emit more strongly at 100~$\mu$m, thereby reducing the radio to FIR emissivity ratio for a given amount of dust. Also the more energetic radiation field may disassociate the VSGs/PAHs responsible for the spinning dust emission. Feature A1, associated with IC~348, has an emissivity similar to an H\textsc{ii} region although no H\textsc{ii} region with a flux density greater than 5~mJy at radio frequencies is seen with the VSA at the position of IC348. Nevertheless, there is a significant UV radiation field arising from the low mass stars comprising the open cluster in the IC348 reflection nebulosity and there may additionally be some UV illumination from the nearby Perseus~OB2 association. The overall radiation field in IC~348 is 10~--~100 times that of the general ISM ~\citep{Bachiller:87}. The higher dust temperature (T$_{dust}$~$\approx$~40~K) compared with that (T$_{dust}$~$\approx$~30~K) in the cooler features confirms the presence of this higher radiation field in the A1 region. The lack of a detectable H\textsc{ii} region indicates a less intense radiation field or a lower gas density than in a normal H\textsc{ii} region. 

Ignoring feature A1, with its lower dust emissivity, the mean dust emissivity for features A2, A3, B and C~(13.8~$\pm$~2.2) is consistent with the integrated value calculated for the whole region by~\citet{Watson:05}. These comparisons of dust emissivity confirm that our detection of excess emission is in agreement with previous observations, and can be explained in terms of spinning dust.

\begin{figure}
\begin{center}
\includegraphics*[scale=0.35,angle=180,viewport=730 80 30 550]{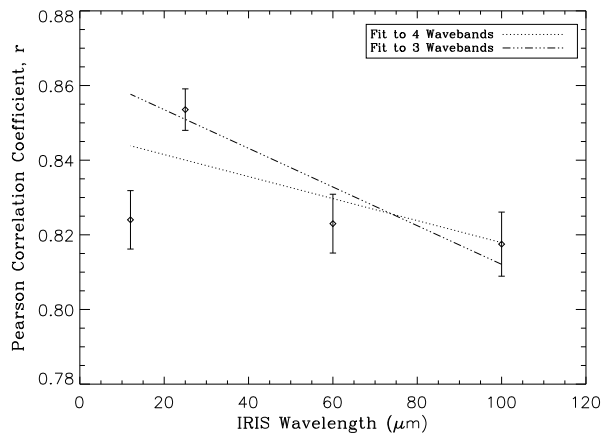}
\caption{Pearson correlation coefficient plotted as a function of wavelength. A fit to all 4 \textit{IRIS} wavebands (dotted line), and a fit to the 25, 60 and 100~$\mu$m wavebands (dash-dot-dot line) are plotted. Both fits show a significant non zero gradient, at a 2.5$\sigma$ and 3.8$\sigma$ level respectively, with the correlation increasing as the wavelength decreases. 1$\sigma$ errorbars are plotted.}
\label{Fig:Correlation}
\end{center}
\end{figure}


\subsection{Correlation between the Microwave and IR Observations}
\label{sec:correlations}

A visual comparison of the \textit{IRIS} images overlaid with the VSA contours in Fig.~\ref{Fig:IRIS_image} suggests a possible correlation between the microwave and IR emission. A Pearson correlation analysis was performed, and Fig.~\ref{Fig:Correlation} illustrates how good the overall correlation is~($\approx$~0.8), and displays the calculated Pearson correlation coefficient as a function of \textit{IRIS} wavelength. The plotted error bars are~1$\sigma$, and were determined using the Fisher z$^{\prime}$ transformation, which converts the non-normal Pearson sampling distribution to a normal distribution with standard errors.

Also plotted on Fig.~\ref{Fig:Correlation} are 2 fits: one fit applied to all 4 \textit{IRIS} wavelengths~(12, 25, 60 and 100~$\mu$m), and one fit applied to only 3 of the \textit{IRIS} wavelengths~(25, 60 and 100~$\mu$m). Both fits display a non-zero gradient at a 2.5$\sigma$ level and a 3.8$\sigma$ level respectively. This implies that the correlation between the microwave and the IR increases as the wavelength decreases. The shorter IR wavelengths are known to be due to emission from the smallest dust grains  (VSGs and PAHs), and hence the fact that the correlation is stronger with shorter wavelengths reinforces the hypothesis of this anomalous emission being caused by spinning dust.

This correlation analysis was performed across 4 of the 5 features in the VSA image (A2, A3, B and C). Feature A1 was excluded because it has different physical conditions compared to the other 4 features as shown in Section~\ref{sec:emiss} and Table~\ref{Table:emiss}. When feature A1 is excluded from the correlation analysis, the overall correlation coefficient increases by~$\approx$~4~\%.

The 12~$\mu$m data point in Fig.~\ref{Fig:Correlation} appears to be low, however we must remember that although the \textit{IRIS} data are an improved version of the original \textit{IRAS} data, there is still some substantial ``contamination'' from point sources in these images. These point sources would be expected to dominate at the shorter wavelengths and hence this could explain the apparent decrease in correlation at 12~$\mu$m. 

Further investigation using data from the \textit{Spitzer Space Telescope} will be performed in a follow-up paper. This should clear up our understanding of whether the point source contamination is responsible for the apparent peak at 25~$\mu$m.


\subsection{Anomalous Emission and Angular Scale}
\label{sec:ang_scale}

G159.6--18.5 was observed by~\citet{Watson:05} with the COSMOSOMAS experiment on angular scales of ~$\approx$~1$^{o}$, and hence could not resolve any of the 5 individual features observed with the VSA in this present work. Using both results allow us to investigate how the anomalous emission varies with angular scale. The total \textit{WMAP} flux density measured by~\citet{Watson:05} was 40.3~$\pm$~0.4~Jy at 33~GHz, and the total flux density in the 5 features in the VSA map was found to be 4.4~$\pm$~0.4~Jy~(Table~\ref{Table:Flux}). This implies that the VSA is only observing a small fraction ($\approx$~10~\%) of the anomalous emission observed by COSMOSOMAS, and that the bulk of the anomalous emission is originating from some large scale structure that is being resolved out by the VSA, due to the short-spacing problem. This idea also explains why the integrated flux density in the filtered VSA image, 0.96~$\pm$~0.12~Jy~(Table~\ref{Table:GB6}) is much lower than the flux density in the unfiltered VSA image. Therefore, the features observed by the VSA must be ``peaks'' of anomalous emission, located on a large scale, diffuse cloud responsible for the~33~GHz emission in\textit{WMAP}.


\section{Conclusions}
\label{sec:con}

Observations of G159.6--18.5 with the VSA in its super-extended array configuration with a synthesized beam of~$\approx$~7$^{\prime}$ FWHM provide clear evidence for excess emission at 33~GHz in 5 distinct dust-correlated components. This anomalous emission cannot be explained by either free--free or thermal dust emission, and as such produces a peaked spectrum. After ruling out both UCH\textsc{ii} regions and GPS by observations of ancillary data, and magnetic dipole emission based on previous observations~\citep{Battistelli:06}, we explained the excess emission by fitting for electric dipole radiation (spinning dust) as predicted by~\citet{DaL:98}.

From the fitted SEDs, feature A1 was found to have~T$_{dust}$~$\approx$~40~K, which was significantly hotter than the other 4 features with~T$_{dust}$~$\approx$~30~K. The central component of the shell peaks at~$\approx$~60~$\mu$m, and as such was found to have~T$_{dust}$~$\approx$~70~K. We suspect this central heating is due to the presence of the central O9.5--B0V star, HD~278942.

The dust emissivity for each of the 5 features were calculated, and found to be in agreement with the values from previous observations where anomalous emission has been detected. Feature A1 was found to have a similar value to that of H\textsc{ii} regions, while the other 4 features all agree with the typical value for high Galactic latitudes.

By performing a correlation analysis between the VSA and \textit{IRIS} observations, we have shown that the correlation is stronger at shorter wavelengths, suggesting that this anomalous emission is caused by the smallest dust grains such as VSGs or PAHs. It is these dust grains, which current spinning dust models~\citep{DaL:98, Ali:09} predict as the source of the electric-dipole radiation, and hence these correlation results imply that what we are observing is emission from spinning dust grains. Spectroscopic observations~\citep{Iglesias:08} have provided evidence for the existence of the naphthalene cation~(a simple PAH) in the G159.6--18.5 region, which could be responsible for the anomalous emission.  

To gain a better understanding of the origin and nature of this anomalous emission, we compared our results with the previous analysis of this region performed by~\citet{Watson:05}. This comparison revealed that the bulk of this anomalous emission ($\approx$~90~\%) appears to be originating from some large scale structure, which is being resolved out by the VSA i.e. that the anomalous emission in G159.6--18.5 is very diffuse and not concentrated in the 5 features observed with the VSA.  

Future high-resolution observations at microwave wavelengths, combined with the \textit{Spitzer} IR data, is needed to investigate in more detail the nature of the anomalous emission. The~\textit{Planck} and~\textit{Herschel} satellites will provide data in the microwave, sub-mm and FIR regimes which are required to measure the spectrum of anomalous dust in the frequency range~$\approx$~10~--~100~GHz.


\section*{Acknowledgments}

CT acknowledges an STFC studentship.
CD acknowledges an STFC Advanced Fellowship.


\bibliographystyle{mn2e}


\bsp 

\label{lastpage}

\end{document}